\RequirePackage{amsmath}
\documentclass[12pt]{iopart}
\usepackage{graphicx}
\usepackage{rotating}
\usepackage{amsmath}
\usepackage{amsfonts}
\usepackage{amssymb}
\usepackage{enumerate}
\usepackage{longtable}
\usepackage{dcolumn}
\usepackage{bm}

\begin{document}

\title{Spin wave excitations in the antiferromagnetic Heisenberg-Kondo model
for heavy fermions.}
\author{M. Acquarone$^1$ }
\author{C. I. Ventura$^2$ }
\address{$^1$IMEM-CNR and Dipartimento di Fisica, Universit\`{a}
di Parma, 43100 Parma, Italy,} 
\address{$^2$Centro At\'omico Bariloche, 8400 - Bariloche,
Argentina ,  } \eads{\mailto{acquarone@fis.unipr.it},%
\mailto{ventura@cab.cnea.gov.ar}}

\date{\textrm{\today}}

\begin{abstract}

Recent inelastic neutron scattering experiments in CeIn$_{3}$ and CePd$_{2}$%
Si$_{2}$ single crystals measured spin wave excitations at low
temperatures. These two heavy fermion compounds exhibit antiferromagnetic
long-range order, but a strong competition between the
Ruderman-Kittel-Kasuya-Yosida(RKKY) interaction and Kondo effect is
evidenced by their nearly equal N\'eel and Kondo temperatures. Our aim is to
show how magnons such as measured in the antiferromagnetic phase of these Ce
compounds, can be described with a microscopic Heisenberg-Kondo model 
introduced by J.R.Iglesias, C.Lacroix and B.Coqblin, used before for studies
of the non-magnetic phase. The model includes the correlated Ce-$4 f$
electrons hybridized with the conduction band, where we also allow for
correlations, and we consider competing
RKKY\ (Heisenberg-like $J_{H} $) and Kondo ($J_{K}$) antiferromagnetic
couplings. Carrying on a series of unitary transformations, we
perturbatively derive a second-order effective Hamiltonian which, projected
onto the antiferromagnetic electron ground state, describes the spin wave
excitations, renormalized by their interaction with correlated itinerant
electrons. We numerically study how the different parameters of the model
influence the renormalization of the magnons, yielding useful information
for the analysis of inelastic neutron scattering experiments in
antiferromagnetic heavy fermion compounds. We also compare our results with
the available experimental data, finding good agreement with the spin wave
measurements in cubic CeIn$_3$.

\end{abstract}

\maketitle

\submitto{\JPCM}
\date{Received: date / Revised version: date}

%

\section{Introduction}

\label{intro}

The description of heavy fermion compounds is challenging due to the rich
variety of phase diagrams they present, and the anomalous physical
properties which may be found. Among them, appear the Ce and U compounds
which exhibit long-range antiferromagnetism (AF) at low temperatures (for
example, among antiferromagnetic Ce compounds, CeRh$_{2}$Si$_{2}$ exhibits
the highest ordering temperature $T_{N} = 36 K$, with local magnetic ordered
moments of $1.34 -1.42 \mu_{B}$ per Ce, i.e. relatively large, compared to
the full Ce$^{3+}$ free-ion value: 2.54 $\mu_B$).\cite{rhmom} Depending on
the particular compound,\cite{radousky} the antiferromagnetism takes
different forms (magnitude of the local moments varies widely: e.g. 0.001 $%
\mu_{B}$ as in CeRu$_2$Si$_2$ or 0.02 $\mu_{B}$ in UPt$_3$, to 1.55 $\mu_{B}$
as in UCu$_5$; as does the spin configuration: with three-, two- or
one-dimensional AF structures observed). Antiferromagnetism may also appear
competing or coexisting with superconductivity,\cite{steglich,mathur} for
which spin fluctuation-mediated pairing mechanisms are explored.\cite{monod}
Non-Fermi liquid behaviour may appear,\cite{stewart} and quantum criticality
has become a subject of intensive study in these compounds, both
experimentally and theoretically.\cite{mathur,schroeder,kuechler,kitp,
coqblin,coleman} The crossover from the antiferromagnetic state to the
non-magnetic heavy fermion state, which can be tuned by pressure, doping or
magnetic field, is one of the most interesting problems in strongly
correlated $f-$compounds.

The physical properties of these compounds are determined by the strongly
correlated $f-$electrons present ($4f$ in Ce; $5f$ in U) and their
hybridization with the conduction band. The RKKY indirect exchange
interaction between $f-$ local magnetic moments, favouring the establishment
of long-range magnetic order, competes with the screening of these moments
by the conduction electrons, described by the Kondo effect.\cite{kondoeff}
This competition is the subject of the Doniach diagram,\cite{doniach} which
compares the variation of the N\'{e}el and Kondo-impurity temperatures with
increasing antiferromagnetic intrasite exchange coupling $J_{K}$, between
local $f$-moments and conduction electron spins. Compounds with similar
magnetic ordering temperature $T_{N}$ and Kondo temperature $T_{K}$, the
temperature below which magnetic susceptibility saturates indicating
coherent Kondo-singlet formation, are ideally suited to the study of this
RKKY-Kondo competition. In this regard, experiments on CeM$_{2}$Sn$_{2}$
(M=Ni, Ir, Cu, Rh, Pd, and Pt: $4d$ or $5d$ transition metals)\cite%
{beyermann} and CeX$_{2}$Si$_{2}$ (X=Au, Pd, Rh, Ru) \cite{severing} were
undertaken, indicating that departures between theory and experiments
resulted from the use of Kondo impurity relations. The Kondo-lattice model,
instead, consisting of a lattice of local magnetic moments coexisting with a
conduction band, has proved appropriate for the description of many $4f$ and 
$5f$ materials, in particular most Ce (or Yb) compounds, respectively
corresponding to a configuration close to $4f^{1}$ (or $4f^{13}$), where one 
$4f$ electron (or hole) interacts with the conduction electrons.\cite%
{coqblin,coqblin1,coqblin2,coqblin3} In 1997 a revisited Doniach diagram was
introduced, including short-range antiferromagnetic correlations in the
Kondo lattice, in order to improve the description and, in particular, to
account for the observed pressure dependence of T$_{K}$ in CeRh$_{2}$Si$_{2}$%
.\cite{coqblin1} The situation is more complex in Uranium compounds, where U
has a $5f^{n}$ configuration with n=2 or 3, since the $5f$ electrons are
much less localized than the $4f$ electrons of rare earths. Regarding spin
dynamics, it is not clear that Ce and U compounds are intrinsically similar.%
\cite{radousky} In the following we will focus on Ce-compounds, except
otherwise specifically stated.

Experimentally, while the magnetic response due to Kondo spin fluctuations
in the paramagnetic state of heavy fermions is well studied, relatively
little is known about the nature of the magnetic excitations in the \textit{%
ordered} phase of Kondo lattices,\cite{radousky} on which our present study
will focus. Being still an unsolved problem how to describe on equal terms
both the Kondo effect and antiferromagnetism,\cite{rech} here we will focus
on systems with relatively large local magnetic moments and study them
deep inside the antiferromagnetic phase: far from the
antiferromagnetic quantum critical point, 
where spin fluctuations would become more relevant.

A few years ago CePd$_2$Si$_2$\cite{cepd2si2} single crystals were studied
with inelastic neutron scattering: below the antiferromagnetic ordering
temperature strongly dispersive spin wave excitations were found, with an
anisotropic damping, which coexisted with the Kondo-type spin fluctuations
also present above $T_N$. At $T=1.5$K these spin waves were measured along
various BZ paths: they were found to present an energy gap of 0.83 meV and
to extend up to almost 3.5 meV. CePd$_2$Si$_2$ has a bcc tetragonal
structure, and its antiferromagnetic ground state is characterized by
propagation vector: $\vec{q}= (1/2, 1/2, 0)$, with ordered moments: $S=0.66
\mu_{B}$, $T_N=8.5$K and $T_{K}=10$K, and linear electronic specific heat
coefficient $\gamma = 250$ mJ$/$mol K$^2$. Under pressure application, at 28.6
kbar the system undergoes a transition into a superconducting phase with
critical temperature of 430 mK.\cite{cein3,llois} More recently, inelastic
neutron studies of CeIn$_3$ single crystals were performed, with similar
results.\cite{cein3} Well defined spin wave excitations with a bandwidth of
2 meV and a gap of 1.28 meV were found in the antiferromagnetic phase,\cite%
{cein3} coexisting with Kondo-type spin fluctuations and crystal-field
excitations which also appeared above $T_N = 10$K$= T_K$. CeIn$_3$
crystallizes in a cubic (fcc) structure, with an antiferromagnetic structure
characterized by magnetic propagation vector $\vec{q}=(1/2, 1/2, 1/2)$, with
ordered moments: $S=0.5 \mu_{B}$ and $\gamma = 130$ mJ$/$mol K$^2$. Under
application of pressure, at 26.5 kbar the system undergoes a transition into
a superconducting phase with critical temperature of 200
mK.\cite{cein3}
 Recently, similar antiferromagnetic magnon excitations were measured 
by inelastic neutron scattering in CeCu$_{2}$, \cite{cecu2}
an anisotropic antiferromagnetic heavy fermion compound with  $T_N= 3.5$K and $T_{K}=4$K.

In next section, we will briefly introduce the microscopic Heisenberg-Kondo
model proposed by J.R.Iglesias, C.Lacroix and B.Coqblin,\cite{coqblin1} to
study the non-magnetic phase of heavy fermion AF compounds, to which we
shall add conduction electron correlations. We will then present our
calculation for the renormalization of spin wave excitations due to their
interaction with the correlated conduction electrons (the Appendix
complements this section). In Section \ref{results}, we will discuss the
results of our study, show how the different parameters of the model
influence the renormalization of the magnons, and compare our results with
the available experimental data.\cite{cepd2si2,cein3} In Section \ref%
{conclusions} we summarize and point out that the present work should yield
useful information for the analysis and prediction of inelastic neutron
scattering experiments in heavy fermion AF compounds, as CeRh$_2$Si$_2$.

\section{Microscopic model, and perturbative approach.}

\label{model1}


In order to describe the Ce-heavy-fermion systems exhibiting
antiferromagnetic long-range order we have used the microscopic model which
has been proposed by Coqblin et al.\cite{coqblin1} to describe the
competition between the Kondo effect and the RKKY interaction, in compounds
where departures from the original Doniach picture\cite{doniach} appear. In
principle, both the RKKY magnetic coupling and the Kondo effect can be
obtained from the Kondo intrasite-exchange term, but when dealing with
approximations it is difficult to assure that both effects are taken into
account if an explicit intersite exchange (as the effective RKKY interaction
or, depending on the system, also the direct exchange) is not included in
the Hamiltonian.\cite{coqblin} The model\cite{coqblin1}consists of a Kondo
lattice, featuring local magnetic moments coupled both to conduction
electrons, by a Kondo-type interaction $J_{K}^{{}}$\ , and, among
themselves, by an antiferromagnetic RKKY-type exchange $J_{H}^{{}}>0$. The
moments are assumed to order below the N\'{e}el temperature $T_{N}$, but we
will concentrate on the zero-temperature limit$.$ The Hubbard-correlated
conduction electrons occupy a non-degenerate band. Therefore, the model may
be represented in standard notation by the following Heisenberg-Kondo
Hamiltonian: \ 
\begin{align}
H& =H_{Heis}+H_{band}^{{}}+H_{Kondo} \\
H_{Heis}& =J_{H}^{{}}\sum_{i\langle m\rangle }\mathbf{S}_{i}\mathbf{\cdot S}%
_{m}-B_{a}\left( \sum_{l\in A}S_{l}^{z}-\sum_{j\in B}S_{j}^{z}\right) \\
H_{band}^{{}}& =\sum_{i\sigma }\epsilon _{i}n_{i\sigma }^{{}}+\
\sum_{i\langle m\rangle \sigma }t_{im}^{{}}c_{i\sigma }^{\dagger }c_{m\sigma
}^{{}}+U\sum_{i}n_{i\uparrow }^{{}}n_{i\downarrow }^{{}} \\
H_{Kondo}& =J_{K}^{{}}\left( \sum_{l\in \mathcal{A}}\mathbf{s}_{l}^{{}}\cdot 
\mathbf{S}_{l}^{{}}+\sum_{j\in \mathcal{B}}\mathbf{s}_{j}^{{}}\cdot \mathbf{S%
}_{j}^{{}}\right)  \label{sp1}
\end{align}

\noindent where \ we have included in $H_{Heis}$ an anisotropy field $%
B_{a}>0.$ which is physically realized by the crystal field in the heavy
fermion systems. It will be shown to play a crucial role with respect to the
stability of the AF spin waves.

Here $i\left\langle m\right\rangle $ $\ $indicates that the lattice site
index $m$\ runs over the $z$\ nearest neighbours of site $i$ which, in turn,
runs over the $N$ sites of the full lattice. The itinerant electron spin is $%
s_{l\left( j\right) }^{{}}$, while $\mathcal{A},\mathcal{B}$ label the two
interpenetrating sublattices with $N/2$ sites each, where local moments, $%
S_{l\in \mathcal{A}}^{{}}$ and $S_{j\in \mathcal{B}}^{{}}$\ , with opposite
moment direction, sit. The Kondo exchange $J_{K}$\ could in principle have
either sign (though for heavy fermion compounds, it would be
antiferromagnetic: $J_{K}\geq 0$).\textit{\ }

\subsection{Diagonalization of $H_{Heis}$.}

We diagonalize the Heisenberg term, by representing the local moments
operators in the Holstein- Primakoff approximation, which is appropriate at
temperatures much lower than $T_{N}$\ . Namely, we take 
\begin{align}
\vec{S}_{l}& =(S_{l}^{+},S_{l}^{-},S_{l}^{z})  \notag \\
& \sim (\sqrt{2S}b_{l},\sqrt{2S}b_{l}^{\dagger },S-b_{l}^{\dagger }b_{l}) 
\notag \\
\vec{S}_{j}& \sim (\sqrt{2S}b_{j}^{\dagger },\sqrt{2S}b_{j}^{{}},-(S-b_{j}^{%
\dagger }b_{j})\,,
\end{align}%
where $b_{l}^{\left( \dagger \right) }$ and $b_{j}^{\left( \dagger \right) }$
are the bosonic spin-deviation operators in sublattices $\mathcal{A}$ and $%
\mathcal{B}$, respectively. \ Going to reciprocal space in the reduced
Brillouin zone (RBZ) one gets an anharmonic Hamiltonian in the
spin-deviation $\left\{ b_{q}^{\dagger },b_{q}^{{}}\right\} $ operators.
It's convenient for what follows to recall that a generic anharmonic
Hamiltonian of the form: 
\begin{eqnarray}
H_{anh}=\sum_{q}F_{q}\left( b_{q}^{\dagger }b_{q}^{{}}+b_{-q}^{\dagger
}b_{-q}^{{}}\right) +\sum_{q}G_{q}\left( b_{q}^{\dagger }b_{-q}^{\dagger
}+b_{q}^{{}}b_{-q}^{{}}\right)
\end{eqnarray}%
where $F_{q},G_{q}$ are $c$-numbers, is diagonalized by a Bogolyubov
transformation, which introduces the AF spin wave operators $\left\{
a_{q}^{\dagger },a_{q}^{{}}\right\} $ according to:%
\begin{align}
b_{q}^{\dagger }& =a_{q}^{\dagger }\mathrm{Ch}\left( \vartheta
_{q}^{{}}\right) +a_{-q}^{{}}\mathrm{Sh}\left( \vartheta _{q}^{{}}\right) 
\notag \\
b_{q}^{{}}& =a_{-q}^{\dagger }\mathrm{Sh}\left( \vartheta _{q}^{{}}\right)
+a_{q}^{{}}\mathrm{Ch}\left( \vartheta _{q}^{{}}\right)
\end{align}

The diagonalization condition is $\ \mathrm{Th}\left( 2\vartheta
_{q}^{{}}\right) =-G_{q}/F_{q}$ . The diagonalized Hamiltonian (defining $\ 
\mathrm{sgn}\left( x\right) =x/\left\vert x\right\vert $ ) reads: 
\begin{eqnarray}
e^{S}H_{anh}e^{-S}=H_{diag}=\sum_{q}\mathrm{sgn}\left( F_{q}\right) \sqrt{%
F_{q}^{2}-G_{q}^{2}}\left( a_{-q}^{\dagger }a_{-q}^{{}}+a_{q}^{\dagger
}a_{q}^{{}}\right)  \label{squeezing general}
\end{eqnarray}

Stability of the system requires the renormalized frequency to be real and
positive. Reality imposes the condition $F_{q}^{2}>G_{q}^{2}$, i.e. the
anharmonic part must have an amplitude smaller than the harmonic part. If
the renormalized frequencies are real, their positiveness is assured by the
additional constraint $F_{q}>0$.

In our case the diagonalization condition for $H_{Heis}$\ \ reads: 
\begin{eqnarray}
\mathrm{Th}\left( 2\vartheta _{q}\right) =-\frac{G_{q}}{F_{q}}=-\left( \frac{%
J_{H}^{{}}S}{zJ_{H}^{{}}S+B_{a}^{{}}}\right) \sum_{\Delta _{lj}}\cos \left(
q\Delta _{lj}\right)  
\label{tanh(2 theta)}
\end{eqnarray}

In the absence of the interaction with the fermions, i.e. in the limit of
vanishing $J_{K}$, the frequency of the bare AF spin waves would be ($z$ is
the number of nearest neighbors and $\Delta _{lj}$ is the vector joining two
n.n.sites)%
\begin{align}
H_{Heis}& =\sum_{q}\hbar \Omega _{q}^{{}}\left( a_{q}^{\dagger }a_{q}^{{}}+%
\frac{1}{2}\right)  \label{bare_H_Heis} \\
\hbar \Omega _{q}^{{}}& =\left( zJ_{H}^{{}}S+B_{a}^{{}}\right) \sqrt{1-\left[
\frac{J_{H}^{{}}S\sum_{\Delta _{lj}}\cos \left( q\Delta _{lj}\right) }{%
zJ_{H}^{{}}S+B_{a}^{{}}}\right] ^{2}}  \label{bare_freq}
\end{align}

\subsection{Diagonalization of $H_{band}$.}

The band electrons described by $H_{band}$ \ can be in either the
paramagnetic (PM) or AF state, according to the values \ of the bandwidth $%
W=2zt$ and of the Hubbard correlation $U$ . To diagonalize $H_{band}$ we
will use a reformulation of Gutzwiller's variational approach for the
description of antiferromagnetism in narrow bands due to Spa\l ek et al.\cite%
{SpalekAF} This approach allows to connect smoothly the PM state for $U\ll W$
to standard (mean-field) Slater band-insulator for $U\approx W$ and to the
localized Mott antiferromagnetic insulator for $U\gg W$. One expresses the
correlation-induced bandwidth reduction in the paramagnetic (PM) state by a
Gutzwiller-type factor $\Phi (n,\rho )$ depending on the band filling $n$
and on the probability of double occupancy $\rho =N^{-1}\sum_{l}\left\langle
n_{l\uparrow }^{{}}n_{l\downarrow }^{{}}\right\rangle .$ The correlated band
energies $\varepsilon _{k}^{U}$ are given by: 
\begin{align}
\varepsilon _{k}^{U}& =\Phi (n,\rho )\varepsilon _{k}^{0}\mathrm{\quad }
\label{GutzwillerPM} \\
\Phi (n,\rho )& =1-\left( \frac{n}{2-n}\right) \left( 1-\frac{4\rho }{n^{2}}%
\right)  \notag
\end{align}%
where $\varepsilon _{k}^{0}$ are the uncorrelated band energies. The $U$-
depending optimal value of $\rho $ is found at zero temperature by
minimizing the PM energy 
\begin{eqnarray}
E^{PM}=\Phi (n,\rho )\sum_{k\sigma }\varepsilon _{k}^{0}\left\langle
n_{k\sigma }^{{}}\right\rangle +NU\rho  \label{Gutzwiller energy}
\end{eqnarray}%
at given $n$ and $U$ .

Assuming that the electrons have an AF ground state of N\'{e}el-type, 
one adopts the standard Slater formalism 
 only with the PM
energies renormalized according to Eq.\ref{GutzwillerPM}. We therefore
introduce the fermion operators for this AF Slater-type state $\left\{
\alpha _{k\sigma }^{\left( \dagger \right) },\beta _{k\sigma }^{\left(
\dagger \right) }\right\} $ by the transformation 
\begin{align}
c_{k\sigma }^{\dagger }& =\alpha _{k\sigma }^{\dagger }\cos \zeta _{k\sigma
}-\beta _{k\sigma }^{\dagger }\sin \zeta _{k\sigma }  \notag \\
c_{k+\mathbf{\mathcal{Q},}\sigma }^{\dagger }& =\alpha _{k\sigma }^{\dagger
}\sin \zeta _{k\sigma }+\beta _{k\sigma }^{\dagger }\cos \zeta _{k\sigma }
\end{align}%
where $\mathbf{\mathcal{Q=}}(1/2,1/2,1/2)$ in units of $2\pi /a$ is the
wavevector characterizing the AF magnetic state and $k$ is a wavevector
belonging to the reduced Brillouin zone (RBZ) defined by $\left\vert \mathbf{%
k}\right\vert \leq \left\vert \mathbf{\mathcal{Q}}\right\vert $. The
diagonalization condition for $H_{band}$ in a lattice with an inversion
center yields: 
\begin{eqnarray}
\tan \left( 2\zeta _{k\sigma }\right) =-\sigma \frac{U\left\langle
s\right\rangle _{\alpha \beta }}{\Phi \varepsilon _{k}}\equiv \sigma \tan
\left( 2\zeta _{k}\right) ,  \label{tan(zeta)}
\end{eqnarray}%
where the amplitude of the AF order parameter $\left\vert \left\langle
s\right\rangle _{\alpha \beta }\right\vert $ of the itinerant electrons
(staggered magnetization, or band electron polarization), is given by:%
\begin{eqnarray}
\left\vert \left\langle s\right\rangle _{\alpha \beta }\right\vert =\frac{1}{%
2N}\sum_{k\in RBZ,\sigma }\left( \left\langle n_{k\sigma }^{\alpha
}\right\rangle -\left\langle n_{k\sigma }^{\beta }\right\rangle \right) \sin
2\zeta _{k}  \label{def.<s_z>}
\end{eqnarray}%
Let's stress that Eq.\ref{def.<s_z>} does not set the sign of $\left\langle
s\right\rangle _{\alpha \beta }$. When introducing the interaction with the
ordered local moments, its sign will be set according to 
\begin{eqnarray}
\left\langle s\right\rangle _{\alpha \beta }=-\mathrm{sgn}\left(
J_{K}\right) \left\vert \left\langle s\right\rangle _{\alpha \beta
}\right\vert  \label{sign szband}
\end{eqnarray}%
so that a positive $J_{K}$ means an antiparallel orientation with respect to
the local moment staggered magnetization, assumed positive by definition,
and viceversa.

The resulting diagonal bare electron Hamiltonian is: 
\begin{eqnarray}
H_{band}^{AF}=\sum_{k\in RBZ,\sigma }\left[ E_{k\sigma }^{\alpha }n_{k\sigma
}^{\alpha }+E_{k\sigma }^{\beta }n_{k\sigma }^{\beta }\right] -UN\left( 
\frac{n^{2}}{4}-\left\langle s\right\rangle _{\alpha \beta }^{2}\right) ,
\label{bareband}
\end{eqnarray}%
where the bare electron eigenenergies\ (actually spin-independent) read $%
\left( x=\alpha ,\beta \right) $: 
\begin{eqnarray}
E_{k\sigma }^{x}=\frac{1}{2}Un+\left( 1-2\delta _{x\alpha }\right) \sqrt{%
\Phi ^{2}\varepsilon _{k}^{2}+U^{2}\left\langle s\right\rangle _{\alpha
\beta }^{2}},
\end{eqnarray}%
where $\delta _{x\alpha }$ denotes the Kronecker delta. Notice that $\alpha $
is the lower subband. \ The subband filling factors $\left\langle n_{k\sigma
}^{x}\right\rangle =\left\langle n_{k,-\sigma }^{x}\right\rangle =\left[
\exp \left( E_{k\sigma }^{x}-\mu \right) /k_{B}T+1\right] ^{-1}$also depend
on $\left\langle s\right\rangle _{\alpha \beta }$ through $E_{k\sigma }^{x}\ 
$so that Eq.\ref{def.<s_z>} has to be solved self-consistently.

\section{The Kondo coupling.}

Taking into account the orientation of the local-moment magnetization, we
shall distinguish in the Kondo term the longitudinal $H_{K}^{z}$ from the
transverse $H_{K}^{\perp }$ contributions. By taking as positive \ $z$
direction the direction of $S_{l\in \mathcal{A}}^{{}}$\ , they are defined
as: 
\begin{align}
H_{K}^{z}& =J_{K}^{{}}\left( \sum_{l\in \mathcal{A}}s_{l}^{z}S_{l}^{z}+%
\sum_{j\in \mathcal{B}}s_{j}^{z}S_{j}^{z}\right)  \label{H_Kondo_long_0} \\
H_{K}^{\perp }& =\frac{J_{K}^{{}}}{2}\left[ \sum_{l\in \mathcal{A}}\left(
s_{l}^{+}S_{l}^{-}+H.c.\right) +\sum_{j\in \mathcal{B}}\left(
s_{j}^{+}S_{j}^{-}+H.c.\right) \right]  \label{H_Kondo_trans_0}
\end{align}

\subsection{The longitudinal part $H_{K}^{z}$ of the Kondo coupling term.}

\label{model3}

It is convenient to rewrite the longitudinal Kondo term by decomposing it
into the two sublattices contributions. For a given sublattice $\mathcal{X=A}%
,\mathcal{B}$ we have from Eq.\ref{H_Kondo_long_0}: 
\begin{align}
& \sum_{l\in \mathcal{X}}\left( S-b_{l}^{\dagger }b_{l}^{{}}\right) s_{l}^{z}
\notag \\
& =\left( 2\delta _{\mathcal{XA}}-1\right) \left[ \frac{1}{2}S\sum_{\sigma
,l\in \mathcal{X}}\sigma n_{l\sigma }^{{}}-\frac{1}{2}\sum_{\sigma ,l\in 
\mathcal{X}}b_{l}^{\dagger }b_{l}^{{}}\sigma n_{l\sigma }^{{}}\right]  \notag
\\
& =H_{\mathcal{X}1}^{z}+H_{\mathcal{X}2}^{z}\qquad \left( \mathcal{X}=%
\mathcal{A},\mathcal{B}\right)
\end{align}

This expression, rewritten in terms of the operators $\left\{ a_{p}^{\left(
\dagger \right) }\right\} $ and $\left\{ \alpha _{p\sigma }^{\left( \dagger
\right) },\beta _{p\sigma }^{\left( \dagger \right) }\right\} $ which
diagonalize, respectively, $H_{Heis}$ and $H_{band}$ , yields several
contributions. The first one, not containing Bose operators, is:%
\begin{align}
H_{\mathcal{A}1}^{z}+H_{\mathcal{B}1}^{z}& =\frac{J_{K}}{2}S\sum_{p,\sigma }%
\left[ \left( n_{p\sigma }^{\alpha }-n_{p\sigma }^{\beta }\right) \sin
\left( 2\zeta _{p}\right) \right]  \notag \\
& +\frac{J_{K}}{2}S\sum_{p,\sigma }\sigma \cos \left( 2\zeta _{p}\right)
\left( \beta _{p\sigma }^{\dagger }\alpha _{p\sigma }^{{}}+\alpha _{p\sigma
}^{\dagger }\beta _{p\sigma }^{{}}\right)  \label{H^z_A1+H^z_B1}
\end{align}%
Notice that the second term in Eq.\ref{H^z_A1+H^z_B1}\ \ describes a
Kondo-induced hybridization between the bare electrons.

The second contribution $H_{\mathcal{X}2}^{z}$($\mathcal{X=A},\mathcal{B}$ )
describes scattering terms between the spin waves and itinerant electrons.
It reads:

\begin{align}
H_{\mathcal{X}2}^{z} & =-\frac{J_{K}}{2N}\sum_{pqr,\sigma}\alpha_{r\sigma
}^{\dagger}\alpha_{p+r-q,\sigma}^{{}}\left[ \left( 2\delta_{\mathcal{XA}%
}-1\right) \sin\zeta_{r}+\sigma\cos\zeta_{r}\right] \times  \notag \\
& \times\left[ \cos\zeta_{p+r-q}+\left( 2\delta_{\mathcal{XA}}-1\right)
\sigma\sin\zeta_{p+r-q}\right] V_{pq}  \notag \\
& -\frac{J_{K}}{2N}\sum_{pqr,\sigma}\alpha_{r\sigma}^{\dagger}\beta
_{p+r-q,\sigma}^{{}}\left[ \left( 2\delta_{\mathcal{XA}}-1\right) \sin
\zeta_{r}+\sigma\cos\zeta_{r}\right] \times  \notag \\
& \times\left[ -\sigma\sin\zeta_{p+r-q}+\left( 2\delta_{\mathcal{XA}%
}-1\right) \cos\zeta_{p+r-q}\right] V_{pq}  \notag \\
& -\frac{J_{K}}{2N}\sum_{pqr,\sigma}\beta_{r\sigma}^{\dagger}\alpha
_{p+r-q,\sigma}^{{}}\left[ -\sin\zeta_{r}+\left( 2\delta_{\mathcal{XA}%
}-1\right) \sigma\cos\zeta_{r}\right] \times  \notag \\
& \times\left[ \cos\zeta_{p+r-q}+\left( 2\delta_{\mathcal{XA}}-1\right)
\sigma\sin\zeta_{p+r-q}\right] V_{pq}  \notag \\
& -\frac{J_{K}}{2N}\sum_{pqr,\sigma}\beta_{r\sigma}^{\dagger}\beta
_{p+r-q,\sigma}^{{}}\left[ -\sin\zeta_{r}+\left( 2\delta_{\mathcal{XA}%
}-1\right) \sigma\cos\zeta_{r}\right] \times  \notag \\
& \times\left[ -\sigma\sin\zeta_{p+r-q}+\left( 2\delta_{\mathcal{XA}%
}-1\right) \cos\zeta_{p+r-q}\right] V_{pq}  \label{Xz_cal(X)2}
\end{align}

where $V_{pq}$ is a bosonic operator:%
\begin{align}
V_{pq}& =a_{p}^{\dagger }a_{q}^{{}}\mathrm{Ch}\left( \vartheta _{q}\right) 
\mathrm{Ch}\left( \vartheta _{p}\right) +a_{-p}^{{}}a_{q}^{{}}\mathrm{Ch}%
\left( \vartheta _{q}\right) \mathrm{Sh}\left( \vartheta _{p}\right)  \notag
\\
& +a_{p}^{\dagger }a_{-q}^{\dagger }\mathrm{Sh}\left( \vartheta _{q}\right) 
\mathrm{Ch}\left( \vartheta _{p}\right) +\left( a_{-q}^{\dagger
}a_{-p}^{{}}+\delta _{pq}\right) \mathrm{Sh}\left( \vartheta _{q}\right) 
\mathrm{Sh}\left( \vartheta _{p}\right)
\end{align}%
By inserting the identity $1=\delta _{pq}+\left( 1-\delta _{pq}\right) $ in
Eq.\ref{Xz_cal(X)2} we can distinguish between the contributions containing
the diagonal and non-diagonal parts of $V_{pq}$. The term in $H_{\mathcal{X}%
2}^{z}$ containing the diagonal part $V_{pp}\delta _{pq}$, after
symmetrizing with respect to $\pm q,$ can be further decomposed as $H_{%
\mathcal{AB}2,diag}^{z0}+H_{\mathcal{AB}2,diag}^{z}$ where the
boson-independent term 
\begin{align}
H_{\mathcal{AB}2,diag}^{z0}& =-\frac{J_{K}}{2}\sum_{p,\sigma }\left[ \sin
\left( 2\zeta _{p}\right) \right] \left( n_{p\sigma }^{\alpha }-n_{p\sigma
}^{\beta }\right) \left( \frac{2}{N}\right) \sum_{q}\mathrm{Sh}^{2}\left(
\vartheta _{q}\right)  \notag \\
& -\frac{J_{K}}{2}\sum_{p,\sigma }\sigma \left[ \cos \left( 2\zeta
_{p}\right) \right] \left( \beta _{p\sigma }^{\dagger }\alpha _{p\sigma
}^{{}}+\alpha _{p\sigma }^{\dagger }\beta _{p\sigma }^{{}}\right) \left( 
\frac{2}{N}\right) \sum_{q}\mathrm{Sh}^{2}\left( \vartheta _{q}\right)
\label{H^(z0)_AB2_diag}
\end{align}%
contributes, together with $H_{\mathcal{A}1}^{z}+H_{\mathcal{B}1}^{z}$ (Eq.%
\ref{H^z_A1+H^z_B1}), to the electronic Hamiltonian, yielding modified band
energies and an additional inter-subband Kondo-induced hybridization of the
bare AF electrons. The remaining term $H_{\mathcal{AB}2,diag}^{z}$ contains
bosons and reads: 
\begin{align}
H_{\mathcal{AB}2,diag}^{z}& =-\frac{J_{K}}{4}\sum_{p,\sigma }\sin \left(
2\zeta _{p}\right) \left( n_{p\sigma }^{\alpha }-n_{p\sigma }^{\beta
}\right) \left( \frac{2}{N}\right) \sum_{q}\mathrm{Ch}\left( 2\vartheta
_{q}\right) \left( a_{q}^{\dagger }a_{q}^{{}}+a_{-q}^{\dagger
}a_{-q}^{{}}\right)  \notag \\
& +\frac{J_{K}}{4}\sum_{p,\sigma }\left[ -\sigma \cos \left( 2\zeta
_{p}\right) \right] \left( \beta _{p\sigma }^{\dagger }\alpha _{p\sigma
}^{{}}+\alpha _{p\sigma }^{\dagger }\beta _{p\sigma }^{{}}\right) \left( 
\frac{2}{N}\right) \sum_{q}\mathrm{Ch}\left( 2\vartheta _{q}\right) \left(
a_{q}^{\dagger }a_{q}^{{}}+a_{-q}^{\dagger }a_{-q}^{{}}\right)  \notag \\
& +\frac{J_{K}}{4}\sum_{p,\sigma }\left[ -\sin \left( 2\zeta _{p}\right) %
\right] \left( n_{p\sigma }^{\alpha }-n_{p\sigma }^{\beta }\right) \left( 
\frac{2}{N}\right) \sum_{q}\mathrm{Sh}\left( 2\vartheta _{q}\right) \left(
a_{q}^{\dagger }a_{-q}^{\dagger }+a_{q}^{{}}a_{-q}^{{}}\right)  \notag \\
& +\frac{J_{K}}{4}\sum_{pq,\sigma }\left[ -\sigma \cos \left( 2\zeta
_{p}\right) \right] \left( \beta _{p\sigma }^{\dagger }\alpha _{p\sigma
}^{{}}+\alpha _{p\sigma }^{\dagger }\beta _{p\sigma }^{{}}\right) \left( 
\frac{2}{N}\right) \sum_{q}\mathrm{Sh}\left( 2\vartheta _{q}\right) \left(
a_{q}^{\dagger }a_{-q}^{\dagger }+a_{q}^{{}}a_{-q}^{{}}\right)
\label{H^z_AB2_diag}
\end{align}

Finally, the non-diagonal part of $H_{\mathcal{A}2}^{z}+H_{\mathcal{B}2}^{z}$
is: \ 
\begin{align}
H_{\mathcal{AB}2,nondiag}^{z}& =\frac{J_{K}}{2}\left( \frac{2}{N}\right)
\sum_{pqr,\sigma }{\LARGE [}\sin \left( \zeta _{r}+\zeta _{p+r-q}\right)
\left( \beta _{r\sigma }^{\dagger }\beta _{p+r-q,\sigma }^{{}}-\alpha
_{r\sigma }^{\dagger }\alpha _{p+r-q,\sigma }^{{}}\right)  \notag \\
& -\sigma \cos \left( \zeta _{r}+\zeta _{p+r-q}\right) \left( \beta
_{r\sigma }^{\dagger }\alpha _{p+r-q,\sigma }^{{}}+\alpha _{r\sigma
}^{\dagger }\beta _{p+r-q,\sigma }^{{}}\right) {\LARGE ]}V_{pq}^{{}}\left(
1-\delta _{pq}\right)  \label{def.I_1}
\end{align}

\subsection{Effect of $H_{K}^{z}$ on the electronic Hamiltonian.}
\label{effectlongK}

The appearance of the above-mentioned Kondo-induced hybridization terms
between the itinerant electrons (Eqs. \ref{H^z_A1+H^z_B1} and \ref%
{H^(z0)_AB2_diag}) suggests to perform a joint diagonalization of $H_{band}$
and such terms. It is convenient to define the number of AF spin waves at
zero temperature for the Heisenberg Hamiltonian $\mathcal{N}_{SW}^{0H}$,
expressing the zero-point deviation for the local moments when $J_{K}=0$ ,
and the measurable amplitude of the local moment polarization $\left\langle
S_{0}^{z}\right\rangle $ as: 
\begin{eqnarray}
\mathcal{N}_{SW}^{0H}=\left( \frac{2}{N}\right) \sum_{q}\mathrm{Sh}%
^{2}\left( \vartheta _{q}\right) \qquad \left\langle S_{0}^{z}\right\rangle
=S-\mathcal{N}_{SW}^{0H}  \label{cal.(N)^AF_bose}
\end{eqnarray}
where $\vartheta _{q}$ was defined in Eq.~\ref{tanh(2 theta)}.

Thus, the hybrid Hamiltonian to be diagonalized may be written: 
\begin{align}
H_{0}^{el}& =\sum_{p\sigma }\left( E_{p\sigma }^{\alpha }n_{p\sigma
}^{\alpha }+E_{p\sigma }^{\beta }n_{p\sigma }^{\beta }\right) +\frac{%
J_{K}\left\langle S_{0}^{z}\right\rangle }{2}\sum_{p,\sigma }\left[ \sin
\left( 2\zeta _{p}\right) \right] \left( n_{p\sigma }^{\alpha }-n_{p\sigma
}^{\beta }\right)  \notag \\
& +\frac{J_{K}\left\langle S_{0}^{z}\right\rangle }{2}\sum_{p,\sigma }\sigma %
\left[ \cos \left( 2\zeta _{p}\right) \right] \left( \beta _{p\sigma
}^{\dagger }\alpha _{p\sigma }^{{}}+\alpha _{p\sigma }^{\dagger }\beta
_{p\sigma }^{{}}\right)  \label{H_electronic_alpha_beta}
\end{align}%
\ The diagonalization is realized by introducing the hybridized Fermi
operators $A_{p\sigma }^{\left( \dagger \right) },B_{p\sigma }^{\left(
\dagger \right) }$ through the unitary transformation: 
\begin{align}
\alpha _{p\sigma }^{\dagger }& =A_{p\sigma }^{\dagger }\cos \xi _{p\sigma
}+B_{p\sigma }^{\dagger }\sin \xi _{p\sigma }  \notag \\
\beta _{p\sigma }^{\dagger }& =-A_{p\sigma }^{\dagger }\sin \xi _{p\sigma
}+B_{p\sigma }^{\dagger }\cos \xi _{p\sigma }  \label{def.AB_hybridized}
\end{align}%
The diagonalization condition requires:

\begin{eqnarray}
\tan \left( 2\xi _{p\sigma }\right) &=&\sigma \frac{J_{K}\left\langle
S_{0}^{z}\right\rangle \cos \left( 2\zeta _{p}\right) }{E_{p}^{\beta
}-E_{p}^{\alpha }+J_{K}\left\langle S_{0}^{z}\right\rangle \sin \left(
2\zeta _{p}\right) }  \label{def.csiangle2} \\
&=&\sigma \frac{J_{K}\left\langle S_{0}^{z}\right\rangle \left\vert
\varepsilon _{p}^{{}}\right\vert }{\left( E_{p}^{\beta }-E_{p}^{\alpha
}\right) \sqrt{\Phi ^{2}\varepsilon _{p}^{2}+U^{2}\left\langle
s^{z}\right\rangle _{\alpha \beta }^{2}}+U\left\vert J_{K}\left\langle
s^{z}\right\rangle _{\alpha \beta }\right\vert \left\langle
S_{0}^{z}\right\rangle }  \notag
\end{eqnarray}%
where in the second line we have explicitated $\sin \left( 2\zeta
_{p}\right) $ and $\cos \left( 2\zeta _{p}\right) .$

Therefore the electronic Hamiltonian in diagonal form reads:%
\begin{eqnarray}
H_{0}^{el}=\sum_{p\sigma }\left( \mathcal{E}_{p\sigma }^{A}n_{p\sigma }^{A}+%
\mathcal{E}_{p\sigma }^{B}n_{p\sigma }^{B}\right)  \label{H_0^el}
\end{eqnarray}%
with hybridized energies given by ($X=A,B$): 
\begin{align}
\mathcal{E}_{p\sigma }^{X}& =\frac{1}{2}\left[ E_{p}^{\beta }+E_{p}^{\alpha }%
\right]  \notag \\
& -\left( \delta _{XA}-\frac{1}{2}\right) \sqrt{\left[ E_{p}^{\beta
}-E_{p}^{\alpha }+J_{K}\left\langle S_{0}^{z}\right\rangle \sin \left(
2\zeta _{p}\right) \right] ^{2}+\left[ J_{K}\left\langle
S_{0}^{z}\right\rangle \cos \left( 2\zeta _{p}\right) \right] ^{2}}
\label{hybridized energies}
\end{align}

The evaluation of the band AF order parameters in the hybridized basis, 
as detailed in the Appendix, yields:%
\begin{eqnarray}
\left\langle s^{z}\right\rangle _{AB}=-\mathrm{sgn}\left( J_{K}\right) \frac{%
1}{2N}\sum_{p,\sigma }\left\{ \sin \left[ 2\left( \zeta _{p}^{{}}+\xi
_{p}^{{}}\right) \right] \left\langle n_{p\sigma }^{A}\right\rangle -\sin %
\left[ 2\left( \zeta _{p}^{{}}-\xi _{p}^{{}}\right) \right] \left\langle
n_{\sigma }^{B}\right\rangle \right\}
\label{band AF moment in AB basis averaged}
\end{eqnarray}

Notice that in the limit $J_{K}\rightarrow 0$ , i.e. $\xi
_{p}^{{}}\rightarrow 0$\ \ we recover the result of Eq.\ref{def.<s_z>} for
the isolated band 
\begin{eqnarray}
\lim_{J_{K}\rightarrow 0}\left\langle s^{z}\right\rangle _{AB}=-\mathrm{sgn}%
\left( J_{K}\right) \frac{1}{2N}\sum_{p,\sigma }\sin \left( 2\zeta
_{p}^{{}}\right) \left[ \left\langle n_{p\sigma }^{\alpha }\right\rangle
-\left\langle n_{p\sigma }^{\beta }\right\rangle \right]
\label{band AF magnetization Slater only}
\end{eqnarray}

Conversely, for the case of a band, too weakly correlated to order
antiferromagnetically by itself, i.e. for $U\rightarrow 0$ , $\zeta
_{p}^{{}}\rightarrow 0$\ ,\ we get the Kondo-induced band staggered moment%
\begin{eqnarray}
\lim_{U\rightarrow 0}\left\langle s^{z}\right\rangle _{AB}=-\mathrm{sgn}%
\left( J_{K}\right) \frac{1}{2N}\sum_{p,\sigma }\sin \left( 2\xi
_{p}^{{}}\right) \left[ \left\langle n_{p\sigma }^{A}\right\rangle
+\left\langle n_{\sigma }^{B}\right\rangle \right]
\label{band AF magnetization Kondo only}
\end{eqnarray}
Therefore, due to Eq.~\ref{def.csiangle2},  in this case the band polarization 
is explicitly proportional to the effective field $(- \mid J_{K} \mid  \left\langle 
S_{0}^{z}\right\rangle)$ provided by the local moments, though oppositely 
oriented.

\subsection{The longitudinal Kondo term $H_{K}^{z}$ in the hybrid $\left\{
A_{p\protect\sigma }^{\left( \dagger \right) },B_{p\protect\sigma }^{\left(
\dagger \right) }\right\} $ basis.}

The longitudinal Kondo term obtained above consists of two contributions,
namely $H_{\mathcal{AB}2,diag}^{z}$( Eq.\ref{H^z_AB2_diag}) \ and $H_{%
\mathcal{AB}2,nondiag}^{z}$ (Eq.\ref{def.I_1}) which have to be
explicitated\ in the electronic hybrid basis$\left\{ A_{p\sigma }^{\left(
\dagger \right) },B_{p\sigma }^{\left( \dagger \right) }\right\} $. Defining 
$Z_{p}=\zeta _{p}-\xi _{p}$, we get: 

\begin{align}
I_{d}^{z}& =H_{\mathcal{AB}2,diag}^{z}\left( \alpha _{p\sigma }^{\left(
\dagger \right) },\beta _{p\sigma }^{\left( \dagger \right) }\Longrightarrow
A_{p\sigma }^{\left( \dagger \right) },B_{p\sigma }^{\left( \dagger \right)
}\right)  \notag \\
& =\frac{J_{K}}{2N}\sum_{p,\sigma }\left[ -\sin \left( 2Z_{p}\right) \left(
n_{p\sigma }^{A}-n_{p\sigma }^{B}\right) -\sigma \cos \left( 2Z_{p}\right)
\left( A_{p\sigma }^{\dagger }B_{p\sigma }^{{}}+B_{p\sigma }^{\dagger
}A_{p\sigma }^{{}}\right) \right] \times  \notag \\
& \times \sum_{q}\mathrm{Ch}\left( 2\vartheta _{q}\right) \left(
a_{q}^{\dagger }a_{q}^{{}}+a_{-q}^{\dagger }a_{-q}^{{}}\right)  \notag \\
& +\frac{J_{K}}{2N}\sum_{p,\sigma }\left[ \sin \left( 2Z_{p}\right) \left(
n_{p\sigma }^{A}-n_{p\sigma }^{B}\right) -\sigma \cos \left( 2Z_{p}\right)
\left( A_{p\sigma }^{\dagger }B_{p\sigma }^{{}}+B_{p\sigma }^{\dagger
}A_{p\sigma }^{{}}\right) \right] \times  \notag \\
& \times \sum_{q}\mathrm{Sh}\left( 2\vartheta _{q}\right) \left(
a_{q}^{\dagger }a_{-q}^{\dagger }+a_{q}^{{}}a_{-q}^{{}}\right)
\label{def L in AB basis}
\end{align}

and%
\begin{align}
I_{nd}^{z}& =H_{\mathcal{AB}2,nondiag}^{z}\left( \alpha _{p\sigma }^{\left(
\dagger \right) },\beta _{p\sigma }^{\left( \dagger \right) }\Longrightarrow
A_{p\sigma }^{\left( \dagger \right) },B_{p\sigma }^{\left( \dagger \right)
}\right)  \notag \\
& =\frac{J_{K}}{N}\sum_{pqr,\sigma }{\LARGE [-}\sin \left[ \left(
Z_{r}+Z_{p+r-q}\right) \right] \left( A_{r\sigma }^{\dagger }A_{p+r-q,\sigma
}^{{}}-B_{r\sigma }^{\dagger }B_{p+r-q,\sigma }^{{}}\right)  \notag \\
& -\sigma \cos \left[ \left( Z_{r}+Z_{p+r-q}\right) \right] \left(
A_{r\sigma }^{\dagger }B_{p+r-q,\sigma }^{{}}+B_{r\sigma }^{\dagger
}A_{p+r-q,\sigma }^{{}}\right) {\LARGE ]}V_{pq}\left( 1-\delta _{pq}\right)
\label{def.I1 in AB basis}
\end{align}

\subsection{The transverse Kondo term $H_{K}^{\perp }$\ in the hybrid $%
\left\{ A_{p\protect\sigma }^{\left( \dagger \right) },B_{p\protect\sigma %
}^{\left( \dagger \right) }\right\} $ basis.}

The transverse Kondo term reads: 
\begin{eqnarray}
H_{K}^{\perp}=\frac{1}{2}J_{K}\left[ \sum_{l\in A,\sigma=\pm}s_{l}^{\sigma
}S_{l}^{-\sigma}+\sum_{j\in B,\sigma=\pm}s_{j}^{\sigma}S_{j}^{-\sigma}\right]
\end{eqnarray}

By expressing the local moments in terms of the spin wave operators $%
a_{q}^{\left( \dagger \right) }$, and the electronic part in terms of the
hybrid operators $\left\{ A_{p\sigma }^{\left( \dagger \right) },B_{p\sigma
}^{\left( \dagger \right) }\right\} $, we get:%
\begin{eqnarray}
I^{\perp } &=&H_{K}^{\perp }\left( \alpha _{p\sigma }^{\left( \dagger
\right) },\beta _{p\sigma }^{\left( \dagger \right) }\Longrightarrow
A_{p\sigma }^{\left( \dagger \right) },B_{p\sigma }^{\left( \dagger \right)
}\right) =  \notag \\
&=&\frac{J_{K}}{2}\sqrt{\frac{S}{N}}\sum_{pq\sigma }A_{p\sigma }^{\dagger
}A_{p+q,-\sigma }^{{}}a_{q}^{\dagger }\left[ \mathrm{Ch}\left( \vartheta
_{q}\right) \mathcal{C}_{AA}^{+-}\left( p,q\right) +\mathrm{Sh}\left(
\vartheta _{q}\right) \mathcal{C}_{BB}^{+-}\left( p,q\right) \right] \\
&&+\frac{J_{K}}{2}\sqrt{\frac{S}{N}}\sum_{pq\sigma }A_{p\sigma }^{\dagger
}A_{p+q,-\sigma }^{{}}a_{-q}^{{}}\left[ \mathrm{Sh}\left( \vartheta
_{q}\right) \mathcal{C}_{AA}^{+-}\left( p,q\right) +\mathrm{Ch}\left(
\vartheta _{q}\right) \mathcal{C}_{BB}^{+-}\left( p,q\right) \right]  \notag
\\
&&+\frac{J_{K}}{2}\sqrt{\frac{S}{N}}\sum_{pq\sigma }\sigma A_{p\sigma
}^{\dagger }B_{p+q,-\sigma }^{{}}a_{q}^{\dagger }\left[ \mathrm{Ch}\left(
\vartheta _{q}\right) \mathcal{C}_{AB}^{+-}\left( p,q\right) -\mathrm{Sh}%
\left( \vartheta _{q}\right) \mathcal{C}_{BA}^{+-}\left( p,q\right) \right] 
\notag \\
&&+\frac{J_{K}}{2}\sqrt{\frac{S}{N}}\sum_{pq\sigma }\sigma A_{p\sigma
}^{\dagger }B_{p+q,-\sigma }^{{}}a_{-q}^{{}}\left[ \mathrm{Sh}\left(
\vartheta _{q}\right) \mathcal{C}_{AB}^{+-}\left( p,q\right) -\mathrm{Ch}%
\left( \vartheta _{q}\right) \mathcal{C}_{BA}^{+-}\left( p,q\right) \right] 
\notag \\
&&+\left( A\rightleftarrows B\right)  \label{Kondo transverse a}
\end{eqnarray}%
where the numerical coefficients $\mathcal{C}_{XY}^{+-}\left( p,q\right) $ ($%
X,Y=A,B$\ ) , depending on the angles $\zeta _{p}$ and $\ \xi _{p}$ (Eqs.\ref%
{tan(zeta)} and \ref{def.csiangle2}) are explicitated in the Appendix. 

\section{Perturbative derivation of the effective magnon Hamiltonian.}

We will now describe the perturbative treatment performed to derive the
effective second-order Hamiltonian for magnons. We begin by rearranging the
total Hamiltonian, separating it into a basic part $H_{0}$, and a
perturbation\ $I$. The basic part consists of the bare\ magnon part $%
H_{Heis} $ (Eq. \ref{bare_H_Heis}) plus the diagonalized electron
Hamiltonian $H_{0}^{el}$ (Eq.\ref{H_0^el}), The perturbation includes the
full transverse Kondo coupling term ($I^{\perp }$) and the boson-dependent
part of the longitudinal Kondo Hamiltonian ($I^{z}$), i.e. the terms not
included in $H_{0\text{ }}^{el}$. Explicitly we have: 
\begin{eqnarray}
H &=&H_{0}+I=H_{0}+\left( I^{z}+I^{\perp }\right)  \notag \\
H_{0} &=&H_{Heis}+H_{0}^{el}  \notag  \label{I perp def} \\
I^{z} &=&I_{d}^{z}+I_{nd}^{z}  \label{I long def}
\end{eqnarray}%
where $I_{d}^{z},I_{nd}^{z}$ and $I^{\perp }$ are respectively given by Eqs.%
\ref{def L in AB basis}, \ref{def.I1 in AB basis} and \ref{Kondo transverse
a}.

In the following , the effect of the total perturbation\ $%
I=I_{d}^{z}+I_{nd}^{z}+I^{\perp }$ will be taken into account through a Fr%
\"{o}hlich-type of truncated unitary transformation \cite{wagner}. We
determine the generator $\mathcal{R}$ of the appropriate canonical
transformation by eliminating from the transformed Hamiltonian the first
order term in the perturbation $I$. To this aim, we impose $I+i[\mathcal{R}%
,H_{0}]=0$. Introducing the notation $\mathcal{R\equiv R}_{d}^{z}+\mathcal{R}%
_{nd}^{z}+\mathcal{R}_{{}}^{\perp }$, we decompose this constraint into
three separate equations, which can be solved \cite{tlpaper} yielding: $%
\mathcal{R}_{{}}^{\left( z,\perp \right) }=\lim_{t\rightarrow 0}\frac{i}{%
\hbar }\int_{-\infty }^{t}I_{{}}^{\left( z,\perp \right) }(x)dx.$ Each term
in the perturbation produces a corresponding term in the generator, namely,
from $I_{d}^{z}$ and $I_{nd}^{z}$\ we obtain the "longitudinal" generators $%
\mathcal{R}_{d}^{z}$ and $\mathcal{R}_{nd}^{z}$\ while from $I^{\perp }$\ we
obtain the \bigskip "transverse" generator $\mathcal{R}_{{}}^{\perp }$\ .
The terms $\mathcal{R}_{d}^{z},\mathcal{R}_{nd}^{z}$\ and $\mathcal{R}%
_{{}}^{\perp }$\ \ are detailed in the Appendix. By this procedure,\cite%
{tlpaper}, we obtain the second-order effective Hamiltonian for the \
magnon-conduction electron system as: 
\begin{eqnarray}
H_{eff}=H_{{}}^{0}+\frac{1}{2}\left[ \mathcal{R}_{d}^{z}+\mathcal{R}%
_{nd}^{z}+\mathcal{R}_{{}}^{\perp },I_{d}^{z}+I_{nd}^{z}+I^{\perp }\right] +%
\mathcal{O}\left( I_{d}^{z},I_{nd}^{z},I^{\perp }\right) ^{3}
\label{controlparameter}
\end{eqnarray}%
$\quad $ Let us anticipate here that one finds that the perturbative
parameter which actually controls this expansion, is the ratio $\left\vert
J_{K}/J_{H}\right\vert $ weighed by coefficients depending on the electronic
band structure and filling, whose expresssions are detailed in the Appendix.
These electronic coefficients effectively reduce the magnitude of the
perturbative control parameter from the raw value $\left\vert
J_{K}/J_{H}\right\vert $, leading to a smooth convergence of the
perturbative expansion even when $\mid J_{K}/J_{H}\mid $ is near or exceeds
one. \ This will become clear when we present our numerical results for the
renormalized magnons in the next section.

Finally, we make a projection onto the AF fermion wavefunction to obtain a
second-order effective Hamiltonian for the magnons $H_{SW}^{eff}$. Let us
remark that, when taking the average $\left\langle H_{eff}\right\rangle
_{Fermi}$over the AF\ Fermi wavefunction, we find 
\begin{eqnarray}
\left\langle \left[ \mathcal{R}_{d\left( nd\right) }^{z},I^{\perp}\right]
\right\rangle _{Fermi}=\left\langle \left[ \mathcal{R}^{\perp},I_{d\left(
nd\right) }^{z}\right] \right\rangle _{Fermi}=\left\langle \left[ \mathcal{R}%
_{d}^{z},I_{nd}^{z}\right] \right\rangle _{Fermi}=\left\langle \left[ 
\mathcal{R}_{nd}^{z},I_{d}^{z}\right] \right\rangle _{Fermi}=0
\end{eqnarray}
so that the effective spin wave Hamiltonian has the simpler form%
\begin{align}
H_{SW}^{eff} & \equiv\left\langle H_{eff}\right\rangle _{Fermi}=  \notag \\
& =\sum_{k\sigma}\left( \mathcal{E}_{k\sigma}^{A}\left\langle
n_{k\sigma}^{A}\right\rangle +\mathcal{E}_{k\sigma}^{B}\left\langle
n_{k\sigma}^{B}\right\rangle \right) +\sum_{q}\frac{\hbar\Omega_{q}}{2}%
\left( a_{q}^{\dagger}a_{q}^{{}}+a_{-q}^{\dagger}a_{-q}^{{}}\right)  \notag
\\
& +\frac{1}{2}\left\langle \left[ \mathcal{R}_{d}^{z},I_{d}^{z}\right] +%
\left[ \mathcal{R}_{nd}^{z},I_{nd}^{z}\right] +\left[ \mathcal{R}^{\perp
},I^{\perp}\right] \right\rangle _{Fermi}
\end{align}

where \ 
\begin{eqnarray}
\left\langle n_{k\sigma}^{X}\right\rangle =\left[ \exp\left( \mathcal{E}%
_{k\sigma}^{X}-\mu\right) /k_{B}T+1\right] ^{-1}
\end{eqnarray}

Each one of the perturbative contributions above can be decomposed as a sum
of harmonic and anharmonic terms: thus%
\begin{align}
\frac{1}{2}\left\langle \left[ \mathcal{R}_{d}^{z},I_{d}^{z}\right]
\right\rangle _{Fermi}& =\sum_{q}\mathcal{G}_{q}^{har}\left( a_{q}^{\dagger
}a_{q}^{{}}+a_{-q}^{\dagger }a_{-q}^{{}}\right)  \notag \\
& +\sum_{q}\mathcal{G}_{q}^{anhar}\left( a_{q}^{\dagger }a_{-q}^{\dagger
}+a_{q}^{{}}a_{-q}^{{}}\right)  \label{Eff_Ham_long_diag}
\end{align}%
\begin{align}
\frac{1}{2}\left\langle \left[ \mathcal{R}_{nd}^{z},I_{nd}^{z}\right]
\right\rangle _{Fermi}& =\frac{1}{4}\sum_{q}\hbar \left( \mathcal{D}%
_{q}^{z+}+\mathcal{D}_{q}^{z-}\right) \left( a_{q}^{\dagger
}a_{q}^{{}}+a_{-q}^{\dagger }a_{-q}^{{}}\right)  \notag \\
& +\frac{1}{4}\sum_{q}\hbar \left( \varpi _{q}^{z+}+\varpi _{q}^{z-}\right)
\left( a_{q}^{\dagger }a_{-q}^{\dagger }+a_{q}^{{}}a_{-q}^{{}}\right)
\label{Eff_Ham_long_non_diag}
\end{align}%
and 
\begin{align}
\ \frac{1}{2}\left\langle \left[ \mathcal{R}_{{}}^{\perp },I^{\perp }\right]
\right\rangle _{Fermi}& =\frac{1}{4}\sum_{q}\sum_{X,Y=A,B}\mathcal{T}%
_{q}^{XY}\left( a_{q}^{\dagger }a_{q}^{{}}+a_{-q}^{\dagger
}a_{-q}^{{}}\right)  \notag \\
& +\frac{1}{4}\sum_{q}\sum_{X,Y=A,B}\left( \mathcal{S}_{q}^{XY1}+\mathcal{S}%
_{q}^{XY2}\right) \left( a_{q}^{\dagger }a_{-q}^{\dagger
}+a_{q}^{{}}a_{-q}^{{}}\right)  \label{Eff_Ham_trans_non_diag}
\end{align}

The numerical coefficients entering Eqs.\ref{Eff_Ham_long_diag} - \ref%
{Eff_Ham_trans_non_diag}\ are given by long and complicated expressions,
which we detail in the Appendix.\ Here we just point out that both the
harmonic and the anharmonic parts in $H_{SW}^{eff}$\ have contributions from
both longitudinal and transverse Kondo terms.

By further defining 
\begin{eqnarray}
\hbar \Phi _{q}=\hbar \Phi _{-q}=\mathcal{G}_{q}^{har}+\frac{1}{4}\left( 
\mathcal{T}_{q}^{AA}+\mathcal{T}_{q}^{BB}+\mathcal{T}_{q}^{AB}+\mathcal{T}%
_{q}^{BA}\right) +\frac{1}{4}\left( \hbar \mathcal{D}_{q}^{z+}+\hbar 
\mathcal{D}_{q}^{z-}\right)
\end{eqnarray}%
and 
\begin{eqnarray}
\hbar \Psi _{q}=\hbar \Psi _{-q}=\mathcal{G}_{q}^{anhar}+\frac{1}{4}%
\sum_{X,Y=A,B}\left( \mathcal{S}_{q}^{XY1}+\mathcal{S}_{q}^{XY2}\right) +%
\frac{1}{4}\hbar \left( \varpi _{q}^{z+}+\varpi _{q}^{z-}\right)
\end{eqnarray}

we arrive at the overall effective Hamiltonian: 
\begin{eqnarray}
\left\langle H\right\rangle _{Fermi}=\sum_{q}\hbar\left( \frac{\Omega_{q}}{2}%
+\Phi_{q}\right) \left(
a_{q}^{\dagger}a_{q}^{{}}+a_{-q}^{\dagger}a_{-q}^{{}}\right)
+\sum_{q}\hbar\Psi_{q}\left(
a_{q}^{\dagger}a_{-q}^{\dagger}+a_{q}^{{}}a_{-q}^{{}}\right) +\mathrm{const.}
\label{Eff_Ham_overall_non_diag}
\end{eqnarray}

With one last Bogolyubov transformation\ 
\begin{align}
d_{q}^{\dagger }& =a_{q}^{\dagger }\mathrm{Ch}\left( \eta _{q}^{{}}\right)
+a_{-q}^{{}}\mathrm{Sh}\left( \eta _{q}^{{}}\right)  \notag \\
d_{q}^{{}}& =a_{-q}^{\dagger }\mathrm{Sh}\left( \eta _{q}^{{}}\right)
+a_{q}^{{}}\mathrm{Ch}\left( \eta _{q}^{{}}\right) \ 
\end{align}%
where 
\begin{eqnarray}
\mathrm{Th}\left( 2\eta _{q}^{{}}\right) =-\frac{\hbar \Psi _{q}}{\Omega
_{q}/2+\Phi _{q}}
\end{eqnarray}%
we diagonalize the effective Hamiltonian for the spin excitations (Eq.\ref%
{Eff_Ham_overall_non_diag}), yielding: \ 
\begin{align}
H_{SW}^{eff}& =\sum_{q}\hbar \left[ \left( \frac{\Omega _{q}}{2}+\Phi
_{q}\right) \sqrt{1-\frac{\Psi _{q}^{2}}{\left( \Omega _{q}/2+\Phi
_{q}\right) ^{2}}}\right] \left( d_{q}^{\dagger }d_{q}^{{}}+d_{-q}^{\dagger
}d_{-q}^{{}}\right)  \notag \\
& \equiv \sum_{q} \, \, \, \hbar 
\,  \widetilde{\Omega }_{q} \, \, \, 
 d_{q}^{\dagger }d_{q}^{{}}  \label{H_eff_final}
\end{align}
where:
\begin{eqnarray}
\widetilde{\Omega }_{q} \equiv 
\mathrm{sgn}\left( \Omega _{q}+2\Phi _{q}\right) 
\sqrt{\left( \Omega _{q}+2\Phi _{q}\right) ^{2}-4\Psi _{q}^{2}}
\label{Omega final ren}
\end{eqnarray}
is the renormalized frequency of the antiferromagnetic spin waves. 
In almost all the
 cases investigated numerically we have found that $\Phi _{q}\leq 0$ so that
the overall effect of the interaction of the local moments with the AF band
is, in general, a softening of the spin waves with respect to the
non-interacting case. Hardening for some wave vectors was obtained only for
extremely large values of $\left\vert J_{K}\right\vert \approx t$, outside 
the range of Kondo coupling values estimated for the heavy fermion compounds 
addressed here. Notice
that, in the absence of the effective anisotropy field $B_{a}$ produced by
the crystal field, at $q=0$, due to $\lim_{q\rightarrow 0}\Omega _{q}=0,$
one would get $\widetilde{\Omega }_{q}<0.$ \ A non-vanishing crystal field
thus appears  necessary for the stability of the \ renormalized spin waves. \
Also, our result of Eq.\ref{Omega final ren} suggests that the gap measured
at $q=0$ should not be taken for a direct estimation of $B_{a}$, 
because it also depends  on the value of $\left\vert J_{K}/J_{H}\right
\vert$.

It is interesting to mention that the observed number of Kondo-renormalized
spin waves in the interacting system $\mathcal{N}_{SW}^{0K}$ is: 
\begin{eqnarray}
\left( \frac{2}{N}\right) \sum_{q}\left\langle d_{q}^{\dagger
}d_{q}^{{}}\right\rangle &=&\left( \frac{2}{N}\right) \sum_{q}\left\langle
a_{q}^{\dagger }a_{q}^{{}}\right\rangle \mathrm{Ch}\left( 2\eta _{q}\right)
+\left( \frac{2}{N}\right) \sum_{q}\mathrm{Sh}^{2}\left( \eta _{q}\right)
\label{squeezed number} \\
&=&\left( \frac{2}{N}\right) \sum_{q}\left\langle b_{q}^{\dagger
}b_{q}^{{}}\right\rangle \mathrm{Ch}\left( 2\vartheta _{q}\right) \mathrm{Ch}%
\left( 2\eta _{q}\right)  \notag \\
&&+\frac{1}{2}\left( \frac{2}{N}\right) \sum_{q}\left[ \mathrm{Ch}\left(
2\vartheta _{q}\right) \mathrm{Ch}\left( 2\eta _{q}\right) -1\right]  \notag
\end{eqnarray}%
thus being always larger than the bare magnon value $\mathcal{N}_{SW}^{0H}$
(Eq.\ref{cal.(N)^AF_bose}). At zero temperature we find that the observed
Kondo-renormalized local-moment polarization $\left\langle
S_{K}^{z}\right\rangle $ is:%
\begin{align}
\lim_{T\rightarrow 0}\left\langle S_{K}^{z}\right\rangle & =S-\frac{1}{2}%
\left( \frac{2}{N}\right) \sum_{q}\left[ \mathrm{Ch}\left( 2\vartheta
_{q}\right) \mathrm{Ch}\left( 2\eta _{q}\right) -1\right]  \notag \\
& =S-\frac{1}{2}\left( \frac{2}{N}\right) \sum_{q}\left[ \frac{1}{\sqrt{%
1-\gamma _{q}^{2}}}\frac{\left\vert \Omega _{q}/2+\Phi _{q}\right\vert }{%
\sqrt{\left( \Omega _{q}/2+\Phi _{q}\right) ^{2}-\Psi _{q}^{2}}}-1\right]
\label{cal.(N)^AF_final}
\end{align}%
indicating that the screening of the local moments due to their Kondo  
interaction with the electrons  enhances the zero-point-motion quantum 
fluctuations of the local moments, 
already present in the bare AF Heisenberg case. This allows for a physical 
 interpretation of the softening effect which, as anticipated above,   
we obtain for these renormalized magnons (detailed results in next section). 
The AF arrangement of local moments produces an effective magnon Hamiltonian 
in the form of a harmonic plus an anharmonic term, 
which we reduced above to a simple harmonic oscillator form 
by the final Bogolyubov transformation.  That type of transformation 
entails the increase of the zero-point motion of the effective local moments. 
thus  reducing the effective local magnetic moments 
 with respect to their value in the absence of Kondo coupling. 
The relevant scale for the bare Heisenberg energies 
is determined by  $z J_H S (+ B_{a})$.
In the interacting case, by diagonalization we arrived to an expression 
of the  effective spin-wave Hamiltonian (Eq.~\ref{H_eff_final}) 
representing it as a new Heisenberg-like harmonic Hamiltonian  
in terms of an effective local moment $<S^z_{K}>$,
 reduced with respect to the full local moment $S$. So  
now the relevant energy scale for the renormalized spin waves 
is  $z <S^z_K> J_H (+B_{a} )$  , 
a value lower than in the non-interacting case due to 
the Kondo screening of the local moments. 
In other words, the dynamical screening of the local moments  is naturally 
reflected in a Kondo-induced softening of the magnon energies.

\vspace{1cm}

\section{Results and discussion.}

\label{results}

In previous section, we obtained a formally simple final expression for
the renormalized antiferromagnetic magnons given by 
 Eq.~(\ref{Omega final ren}), 
but one which depends on a series of coefficients 
which are detailed in the Appendix. 
The quite complicated explicit expressions for these perturbative coefficients,
depend on the combined effect or interplay of the different model parameters, 
and most coefficients involve multiple summmations over the reduced
Brillouin zone. Therefore, our renormalized magnon results can only be
evaluated numerically, exploring wide ranges of the different model
parameters, in order to assess and compare the main effect of each of them. 
We will show that our model can reasonably explain experimental magnon results 
\cite{cepd2si2,cein3,cecu2} employing parameters in the range 
which has been independently shown \cite{lobos} 
to be appropriate for a phenomenological fit of specific heat measurements 
in this family of antiferromagnetic heavy fermions. 
Our exploration of wide parameter ranges has also allowed us to
verify that the convergence radius of the perturbative series which
determines the magnon renormalization is much wider than one might naively
have expected. As mentioned below Eq.~(\ref{controlparameter}), the results
presented in this section evidentiate that the actual \textquotedblleft
small parameter\textquotedblright\ controlling this perturbative expansion
is not just the bare $\left\vert J_{K}/J_{H}\right\vert $ ratio. Indeed, in
the renormalized magnon frequency this quantity appears always weighed by
electronic structure- and filling-dependent coefficients which, when
combined with suitable values of the other model parameters, effectively
reduce the control parameter value from this ratio, leading to convergent
results also for $\left\vert J_{K}/J_{H}\right\vert >1$, as will be shown in this section.

The numerical study of our model has been done assuming two simple cubic
interpenetrating magnetic sublattices, for simplicity and without much loss
of generality (rigorously, under this assumption our results would
correspond to a cubic (bcc) chemical lattice). Notice that CeIn$_{3}$%
,\cite{cein3} one of the compounds where inelastic neutron scattering (INS)
on single crystals has measured the magnons we aim to describe, is cubic
(though fcc) and has a three-dimensional N\'{e}el-type antiferromagnetic
structure. We evaluated the magnons at zero temperature, assuming an
underlying 3D N\'{e}el-type antiferromagnetic ground state of the system.
Our study focuses on parameter sets far away from the quantum critical
region of these systems, i.e. deep inside the antiferromagnetic phase. In
fact, our parameters lie well inside the AF stable region recently
determined by a DMFT + NRG study\cite{pruschke} of the magnetic phase
diagram of the correlated Kondo-lattice (corresponding to the $J_{H}=0$ case
of our model: the addition of non-negligible AF-like RKKY coupling $J_{H}$,
as done here, will only increase the stability of the AF phase). As
experimentally observed,\cite{cein3} in this range one might expect
Kondo-type spin fluctuations to be less relevant, and the dispersive spin
waves, object of our study, to appear in the AF phase. 
For simplicity, but also in accordance with experimental indications\cite%
{cein3} , we have further assumed that there is one isolated crystal field
level of Ce$^{3+}$ which is relevant, hosting a spin $S=1/2$ (in fact, $%
S=0.5\mu _{B}$ is the experimental magnitude of the local moments in CeIn$%
_{3}$\cite{cein3} ) . The hopping parameter $t$ was taken as unit of energy,
being $W=12t$ the total bare electron bandwidth.

\begin{figure}[tb]
\begin{picture}(0,0)\includegraphics{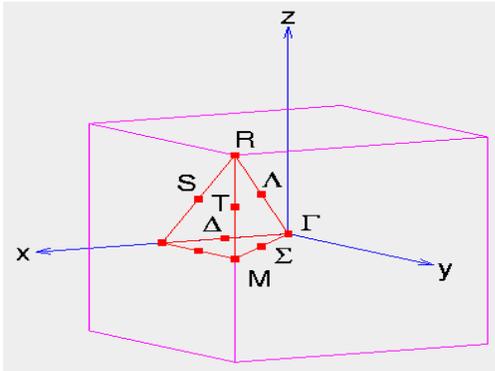}
\end{picture}\setlength{\unitlength}{1243sp}\begingroup\makeatletter\ifx%
\SetFigFont\undefined\gdef\SetFigFont#1#2#3#4#5{\ \reset@font%
\fontsize{#1}{#2pt} \fontfamily{#3}\fontseries{#4}\fontshape{#5} \selectfont}%
\fi\endgroup\begin{picture}(9900,7380)(811,-7261)
\end{picture}
\caption{Simple cubic lattice: 1st Brillouin zone and special symmetry
points.}
\label{crystal}
\end{figure}

We shall start by discussing the general trends we have found in our 
numerical study of the renormalized magnons given by Eq.~\ref{Omega final ren},
to subsequently focus on the description of the measured magnons in
antiferromagnetic heavy fermion compounds. Once one takes into account the
relative orientation of the band and local moment staggered polarizations
according to Eq.\ref{sign szband}, the results of our second-order perturbative treatment 
turn out to be independent of the sign of $J_{K}.$

\begin{figure}[b]
\includegraphics[angle=270 , width=\columnwidth]{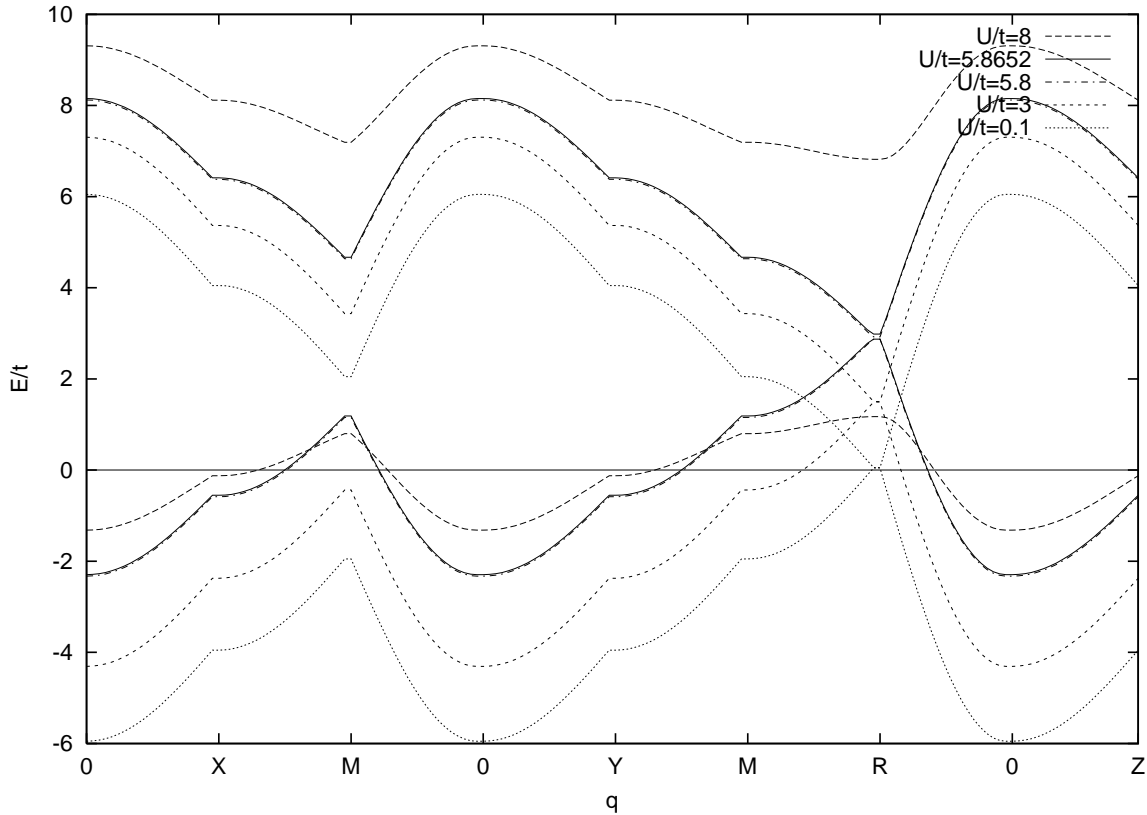} 
\caption{ Hybridized (A,B) electron bands ($\uparrow $) along 
selected BZ paths.
Parameters: $S=1/2; T=0; t=1eV; J_{H}/t=0.001; J_{K}/J_{H}=20; n=0.999$. 
$U/t$ values  as detailed in inset.
}
\label{bands}
\end{figure}

Fig.~\ref{crystal} depicts the first Brillouin Zone (BZ) of the simple
cubic lattice, and we include the notation for the special symmetry points
and BZ paths on which the spin waves were numerically evaluated ($\Gamma
\equiv O$ denotes the zone center, $\Delta \equiv X=0.5\pi /a\,(1,0,0)$, $%
Y=0.5\pi /a\,(0,1,0)$),$Z=0.5\pi /a\,(0,0,1)$ ). Let us mention here that
for the BZ summations we have used the special-points BZ sampling method by
Chadi-Cohen (CC), at 4th order, for the simple cubic lattice.\cite{macot} To
obtain dressed magnons with the correct symmetry of the lattice, and to take
into account that the multiple summations involve linear combinations of
wavectors, we have noticed \cite{tlpaper}\ that it was not sufficient to include
the basic CC\ set of wave vectors of the (reduced) first BZ octant, but one
needs to extend it to the full first Reduced BZ (RBZ). Therefore one has to
consider all the vectors obtained by applying the 48 symmetry operations of
the $O_{h}$ group to the basic CC set. Thus, at 4th-order of the CC method,
we have included 5760 special symmetry points for each RBZ summation. To
achieve higher accuracy for our determination of the Fermi level (a delicate
issue close to half-filling), we have summed over the 5th order
symmetry-extended CC vector set, thus having used 39168 special points.

In Fig.~\ref{bands} we show the typical conduction electron bandstructure
near half-filling, given by Eq.~(\ref{hybridized energies}). Notice that for
each value of $U$ only the lower band (denoted $A$) is filled. Chemical
potential values obtained for the cases shown,  are:
 $\mu /t= -0.63, 0.84, 2.30, 2.33, 1.13$ (the Fermi level is slightly below
the top of the lower subband),  respectively
 for $U/W=0.008,0.25,0.48,0.49, 0.67$ (i.e. the $U/t$ values of Fig.~\ref{bands}. 
Near the two values: $U/t=5.8, 5.8652$, the conduction
electrons  start to develop an AF spin polarization (see polarization values 
reported  in Table I), which at $U/t=8$ has increased to 0.35. 
The AF solution  has a direct band gap, 
determined by the AF hybridized subband energies at the BZ point $R$ 
on the cube diagonal (see Fig.~\ref{crystal}). 
The size of the gap, as well as the energy of its centroid, 
 increases with the magnitude of the electron correlation $U$, 
while also a correlation-driven band-narrowing
effect is seen to appear.
For small $J_{K}$ , the AF band gap value
essentially depends on $U\left\langle s\right\rangle _{AB},$ with $%
\left\langle s\right\rangle _{AB}\approx \left\langle s\right\rangle
_{\alpha \beta }$ growing with both the band filling $n$ and the correlation 
$U.$ In Table 1 these trends of the band polarization with the different
parameters are evidenced.

          It is also interesting to compare our band  
polarization values (tabulated in Table~\ref{table1}) with those reported in
Fig.2 of Ref.~\cite{pruschke}, for corresponding parameters. 
Notwithstanding the different respective treatments for the band electron
correlations, we find quite reasonable agreement where we could check it.
 In the present work, focused on applications to the AF heavy fermion
 compounds where magnons were measured, 
we have used antiferromagnetic Kondo coupling values in a
relatively narrow region around $J_{K}/W\sim 10^{-5}-10^{-3}$. In agreement
with Peters and Pruschke,\cite{pruschke} in this parameter range we find spin
polarizations characteristic of the RKKY regime:
the local moments are almost fully polarized, while the corresponding
 band polarization obtained for $U \sim 0$ (i.e. unpolarized ``PM'' bare 
conduction band, corresponding e.g. to our data of the first two columns of Table I) 
 is proportional to the ``effective field''  provided by the local spins,
as mentioned in the last paragraph  of Section~\ref{effectlongK}. 
Our $U/W= 0$ results  thus agree quite well with the corresponding ones 
of Ref.~\cite{pruschke} Since their results are given only for $U/W= 0, 1$  
we could not make  the comparison  for intermediate $U/W$ values. 

We have also explored wider ranges of values for $J_{K}$, 
to verify the consequences of Eq.\ref{cal.(N)^AF_final}. We have found that, 
if the renormalized spin waves
 are real and positive everywhere in the RBZ, then the local moment value is
scarcely affected (reduced). Conversely, the larger are the regions of the RBZ where
the Eq.\ref{Omega final ren} yields either negative or imaginary results,
the stronger is the reduction of $\left\langle S_{K}^{z}\right\rangle $, in
qualitative agreement with the results of Ref.~\cite{pruschke} for AF\
Kondo coupling.

\begin{figure}[h]
\includegraphics[angle=270 , width=\columnwidth]{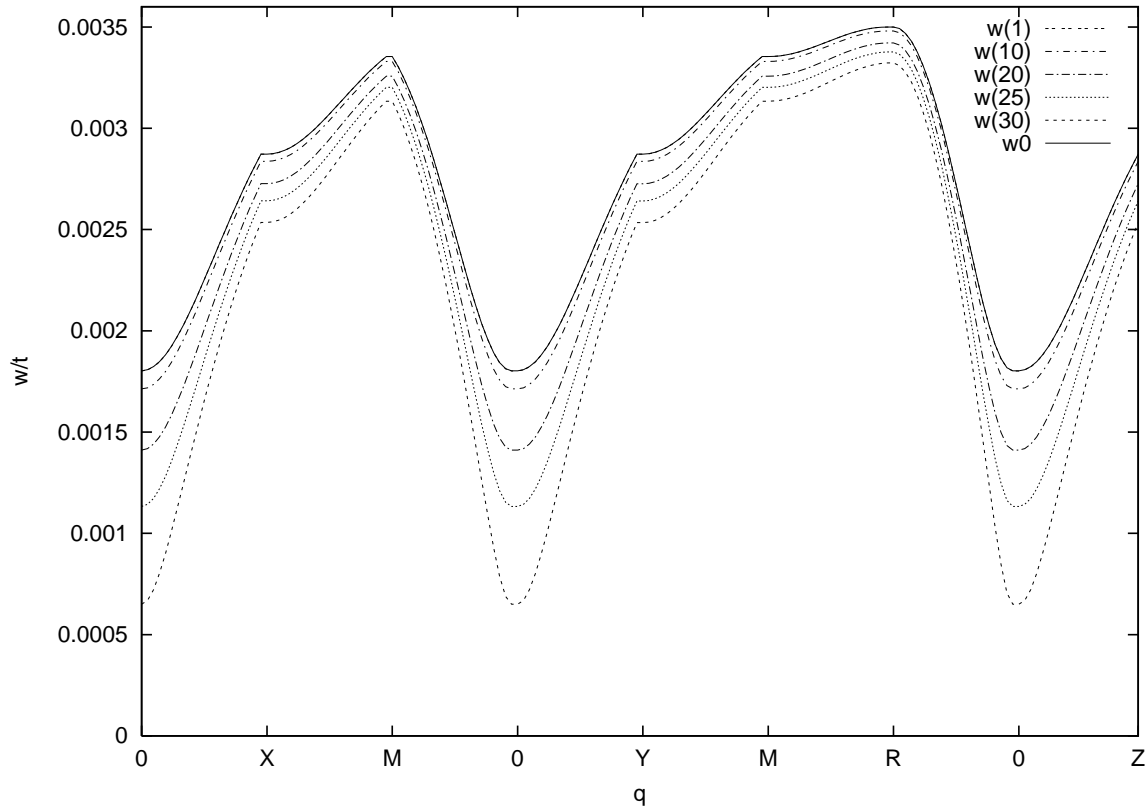} 
\caption{(PM) $J_{K}/J_{H}$ dependence of renormalized AF magnons: magnon
energy along selected BZ paths. Parameters: $
S=1/2;T=0;t=1eV;J_{H}/t=0.001;n=0.999;U/t=6$; $B_{a}/t = 0.0005$.
 $J_{K}/J_{H}$ values as detailed in inset; $w0$ denotes 
the bare magnons $\Omega_{q}$.
}
\label{jratiodepPM}
\end{figure}

\begin{figure}[h]
\includegraphics[angle=270 , width=\columnwidth]{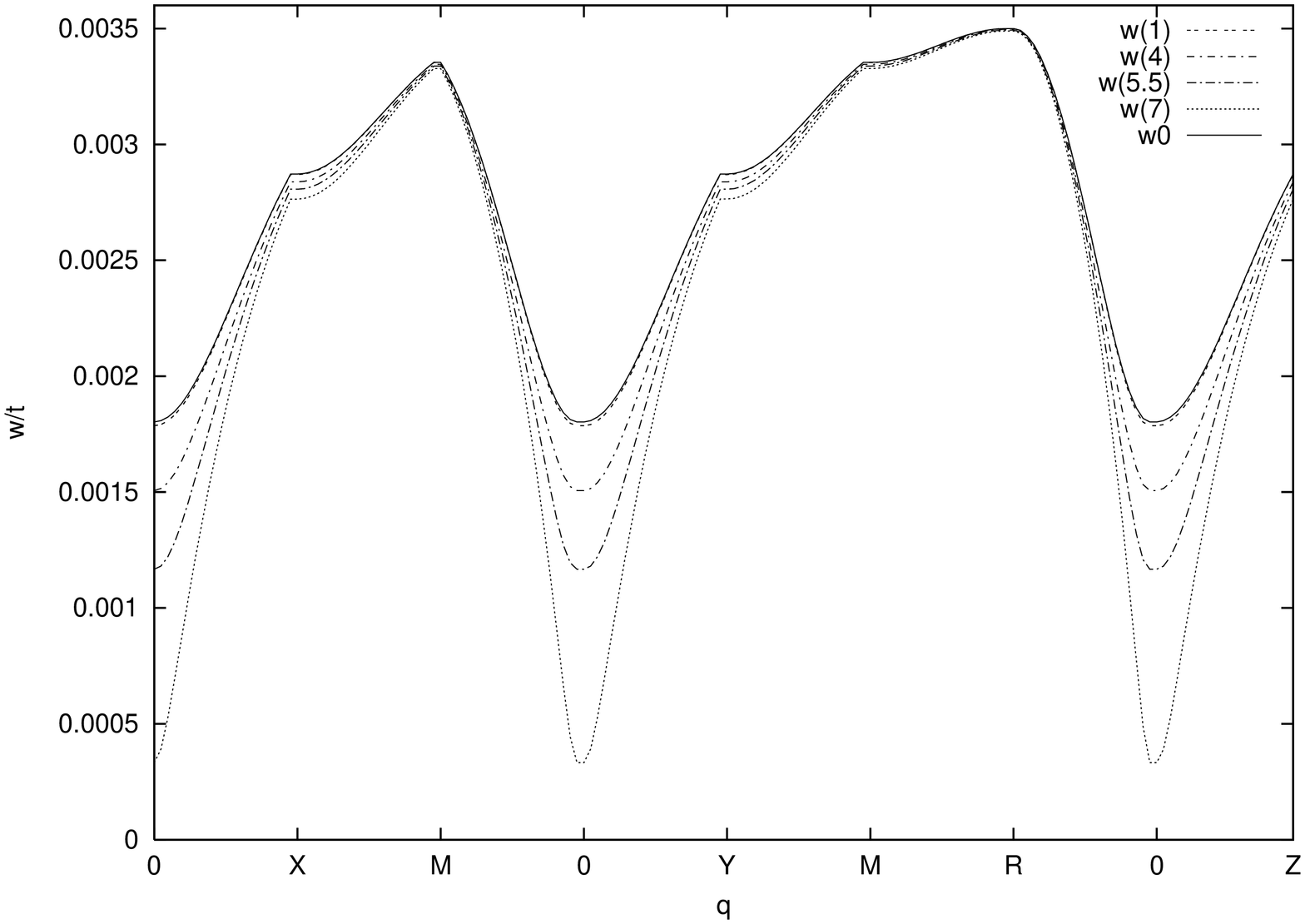} 
\caption{(AF) $J_{K}/J_{H}$ dependence of renormalized AF magnons: magnon
energy along selected BZ paths. Parameters: $
S=1/2;T=0;t=1eV;J_{H}/t=0.001;n=0.999;U/t=6$; $B_{a}/t = 0.0005$.
 $J_{K}/J_{H}$ values as detailed in inset; $w0$ denotes 
the bare magnons $\Omega_{q}$.
}
\label{jratiodepAF}
\end{figure}

In Fig.~\ref{jratiodepPM} we show the dependence of the renormalization 
of the antiferromagnetic 
magnons on the bare $J_{K}/J_{H}$ ratio at half-filling \ ( the most
relevant filling for AF heavy fermion compounds) when the correlation $U=3t$
is not strong enough to appreciably AF-polarize the bare band ($%
U\left\langle s^{z}\right\rangle _{\alpha \beta }\approx 0$ ): we labelled this case 
as PM, corresponding to a paramagnetic bare conduction band (at $J_K = 0$).
The top curve
represents the bare magnons $\Omega _{q}$ (independent of the conduction
electrons). We have included a small anisotropy field $B_{a}$, 
as inelastic neutron scattering experiments reporting a magnon gap
indicate.\cite{cepd2si2,cein3}
 As a general trend, anticipated in the last paragraph of previous section, 
we find that the renormalization reduces the
spin wave frequency $\widetilde{\Omega }_{q}$ with respect to bare 
$\Omega _{q},$ the effect growing 
with the ratio $\left\vert J_{K}/J_{H}\right\vert $ . The
softening is present throughout the whole RBZ, and is strongly dependent on
the wavevector, being maximal around the $\Gamma $ point. In 
Fig.~\ref{jratiodepAF} 
we show the same quantities when the correlation $U=6t$ is
 strong enough to start polarizing antiferromagnetically the  
bare conduction band, see Table I (therefore we have labelled this: AF). 
For higher $U/t$ one finds that the convergence 
of our perturbative series for the renormalized magnons (leading to physical 
 non-negative energies) is limited to  
  a more restricted range of
values of the ratio $\left\vert J_{K}/J_{H}\right\vert $, with respect to the
low correlation case. The reason for this behaviour is that the fermions
which effectively interact with the local moments are those in a
neighbourhood of the Fermi energy of width $\approx \hbar \Omega _{q_{F}}$, as can
be seen by looking at the explicit expressions of the renormalization 
coefficients in the Appendix. We use a Fr\"{o}hlich-type of transformation as in
the BCS\ theory of superconductivity, so that the same type of
considerations about the effective interactions apply. Though not too visible 
in the cases shown  of Fig.~\ref{bands}, which lead to well-defined 
renormalized magnons (except $U/t=8$), 
we have checked which are the main band-structure 
changes at larger $U/t$ values. 
Near half-filling, with the Fermi wavector at $R,$ 
for $U\left\langle s^{z}\right\rangle _{\alpha \beta
}\approx 0$ ($U/t=0.1,3.0)$ the AF subbands disperse strongly 
around $E_{F}$   resulting in a weak interaction with the local moments. 
But as the correlation increases, and 
$U\left\langle s^{z}\right\rangle _{\alpha \beta }$ too, 
the lower AF subband, which contains the Fermi level around $R$, progressively
flattens  so that now more electrons effectively interact with the local moments. 
For the cases $U/t=6, 8$ one also finds that another Fermi surface pocket 
appears around $M.$ 

\begin{figure}[hb]
\includegraphics[angle=270 , width=\columnwidth]{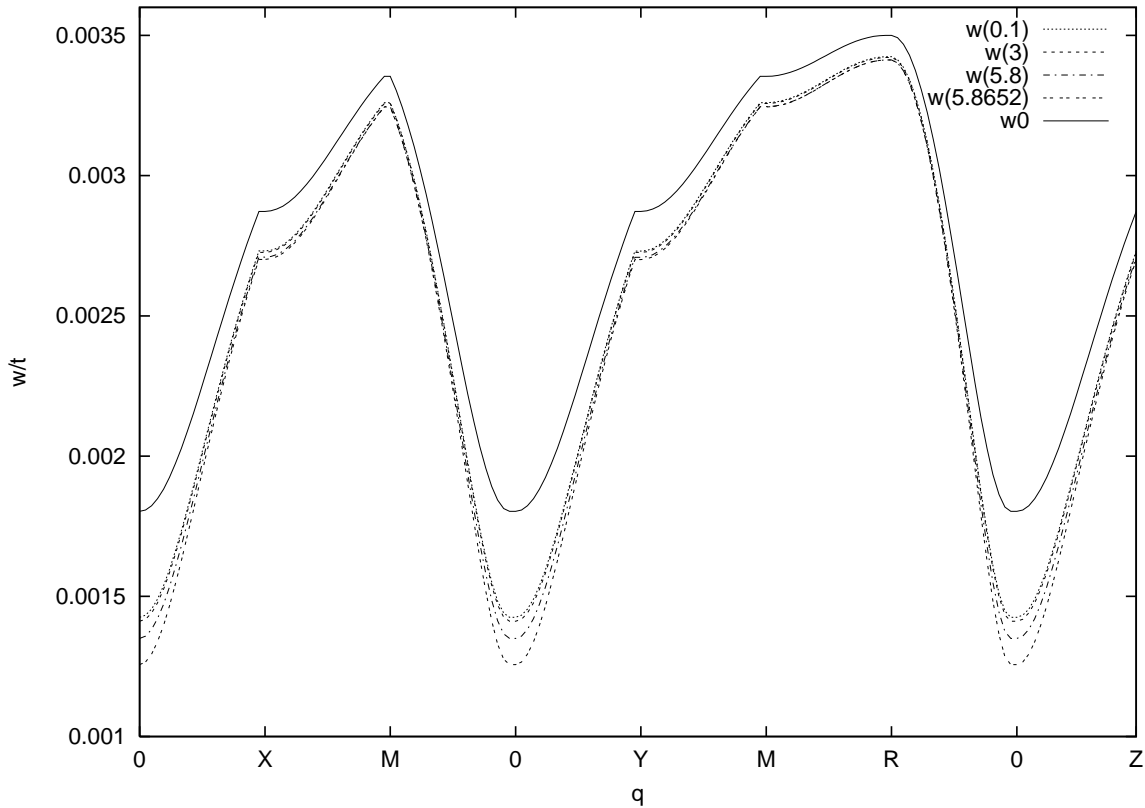} 
\caption{$U$ dependence of renormalized AF magnons: magnon energy along
selected BZ paths. Parameters: $U/t$ as detailed in inset;  $B_{a}/t =
0.0005$; others as in Fig. 2.
}
\label{udep}
\end{figure}

 Fig.~\ref{udep}  depicts the dependence of the renormalized
magnons on the correlations in the conduction band ($U$), at half-filling: 
the trend is a larger softening 
of the magnons when $U$ is increased , and we show
 cases where $U/W$ ranges between 0.008 (for $U/t=0.1$) and 0.49 ($U/t=5.8652$).
Notice that the increase of $U$ increases the $q$-dependence of  the
magnon renormalization. This results from the indirect effect
which $U$ has on magnons (while $J_{K}$ has also a direct effect, since it
appears also as explicit multiplicative factor of the perturbatively
obtained magnon corrections). $U$ affects magnons through the modifications
it induces in electron bandstructure, as discussed above in connection with
Figs.~\ref{jratiodepPM} and \ref{jratiodepAF} , and through the $U$%
-dependence of the energy denominators in the perturbative coefficients
which determine the $q$-dependent renormalization of magnons (details in the
Appendix). The recent more
refined DMFT+NRG treatment of correlations in an extended Kondo lattice model%
\cite{pruschke} unfortunately does not allow us comparison, here, as their
finite U results are presented for antiferromagnetic $J_{K}$ outside the
region of interest in our problem: their correlated AF Kondo coupling system
is studied at much too large Kondo coupling (namely, $J_{K}=0.5W=U$) for the
antiferromagnetic state to remain stable, the stable phase near half filling
in that case being the Kondo insulator with all moments locally quenched. 
\begin{figure}[h]
\includegraphics[angle=270 , width=\columnwidth]{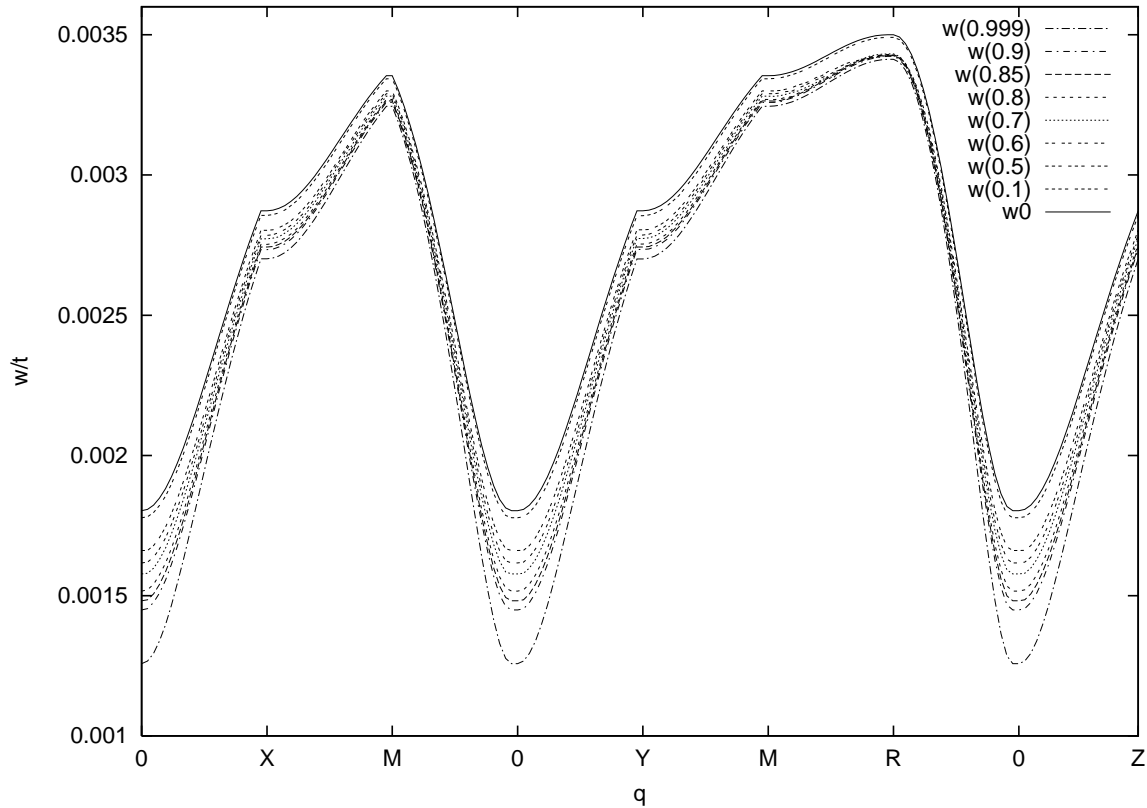} 
\caption{Filling $(n)$ dependence of renormalized AF magnons: magnon energy
along selected BZ paths. Parameters: 
$n$ values as detailed in inset; $U/t=5.865$;  $B_{a}/t = 0.0005$; 
others as in Fig. 2. 
}
\label{ndep}
\end{figure}

We exhibit the effects of doping on the magnon renormalization in Fig.~\ref%
{ndep}. Here the deviation of the renormalized magnon energies from the
bare magnon values increases with the filling: at half-filling the
renormalization is largest, due to more conduction electrons contributing
to the renormalization of magnons by coupling through Kondo interaction to the local
moments. Doping away from half-filling we obtain a smooth reduction of such
renormalization effects, as one would expect. 
Thus, both filling and electron correlation do
increase the renormalization effects, and we have already mentioned that both
lead to similar increases of the spin polarization of the conduction band (see Table I
and Figs. 5-6).
When entering the filling range of values at which the bare electron 
band ($J_K=0$)  develops an itinerant AF polarization, 
the renormalization effects become much stronger, 
as shown by the very different behaviour of the cases $n=0.9$
(PM band for $J_{K}=0$) and $n=0.999$ (AF band for $J_{K}=0$). 

Let us briefly refer again to the $q$-dependence of the AF magnon
renormalization we find. Some anisotropy is present: a larger $q$-dependence
is noticeable along BZ diagonal paths such as $O-M$ or $O-R$ (see e.g. Fig.~%
\ref{ndep}) or paths along the symmetry axes. The renormalization effects
are more pronounced at long wavelenghts: in particular, they are maximal at
the BZ center where we find a spin stiffness decreasing with doping, and
increasing with $U$ or $J_{K}/J_{H}$. Making allowance for the quite
different systems involved, let us mention that the renormalized AF magnon
behavior we obtain contrasts with the one recently disclosed by INS
measurements in \textit{ferromagnetic metallic manganites}:\cite{ye} where
at low-$q$ the spin wave stiffness appears insensitive to doping, while magnons
exhibit a doping-dependent renormalization at the BZ boundaries (recently
suggested to be related to electronic correlations\cite{kapetanakis}).

At this point, let us compare our results with the few INS magnon
measurements available for single crystals of antiferromagnetic heavy
fermions. Comparison in more detail may be made only with CeIn$_{3}$, which
is cubic (though f.c.c.) and presents a three-dimensional AF order as we
have assumed for our calculation. 
The sets of parameters we find as  allowing us a 
reasonable  description of the INS magnon results which were described 
in the Introduction,\cite{cein3} 
as evidenced by inspection of the cases presented in our figures, 
are similar  to the parameter ranges  independently suggested
by other authors for this family of compounds. A concrete example is the
 fit of experimental specific heat curves for CeIn$_{3}$ , made
by Lobos et al.,\cite{lobos} using a model related though not exactly identical   
 to ours,  within a phenomenological approach, who find:
 $J_{H}/t=0.0014$, $t=0.5eV$, $n=1$ 
and $J_{K}/J_{H}=980$. For CeRh$_{2}$Si$_{2}$ they instead estimate: 
$J_{H}/t=0.0034$ and $J_{K}/J_{H}=430$; and their data extrapolation for CePd$%
_{2}$Si$_{2}$ was: $J_{H}/t=0.0034$ with a negligible $J_{K}/J_{H}$.\cite{lobos}

\section{Summary}

\label{conclusions}

In the present work, we have studied spin wave excitations in heavy fermion
compounds with antiferromagnetic long-range order , where a strong
competition of RKKY and Kondo screening is present, as evidenced by nearly
equal magnetic ordering and Kondo temperatures. We have described these
systems using a microscopic model including a lattice of correlated $f$%
-electron orbitals (as in $Ce$-, $U$- compounds of this family) hybridized
with a correlated conduction band, in the presence of competing
RKKY-Heisenberg and Kondo magnetic couplings.

Through a series of unitary transformations we perturbatively derived a
second-order effective Hamiltonian describing the antiferromagnetic 
 spin wave excitations, renormalized by their interaction 
with the conduction electrons. We have
numerically studied the effect of the different parameters of this effective
model on the magnon energy renormalization. Apart from the expected increase 
of renormalization effects for larger Kondo coupling,  we identify another relevant ingredient.
 Magnon renormalization is also amplified  by  spin polarization of the conduction electrons: 
either if it originates from large correlations between the carriers  
or by an increase of the electron filling.       
We have been able to find appropriate sets of model parameters to describe
the few existing measurements of magnons by inelastic neutron scattering in
single crystal samples of antiferromagnetic heavy fermion $Ce$ compounds (such as CeIn$_3$,
CePd$_2$Si$_2$, CeCu$_2$).
Our parameter sets agree with the ranges independently proposed for these
materials, by phenomenological fits of other experiments like specific
heat. Our results may provide information of interest for the prediction of
inelastic neutron scattering experiments in other systems of this family,
like CeRh$_{2}$Si$_{2}$, where there have been suggestions that the RKKY
coupling should be stronger than the Kondo effect,\cite{severing} and the only
existing NIS measurements are of poor quality: they were made on polycrystals%
\cite{severing} many years ago, with lower resolution.

As outlook towards related future work, we might mention the description of
the experimentally reported magnon damping effects, and the study of the
coexistence of antiferromagnetism and superconductivity in the context of
the present model.

\ack
We thank J. Sereni, E. M\"{u}ller-Hartmann, G. Aeppli, P. Coleman, P.
Gegenwart, P. Santini, A. Lobos, J.R. Iglesias and C. Lacroix for discussions 
and references. M.A. thanks Centro At\'{o}mico Bariloche for the hospitality and
support. C.I.V. is Investigador Cient\'{\i}fico of CONICET (Argentina) and
Regular Associate of ICTP (International Centre
for Theoretical Physics, Trieste), and
acknowledges support from CONICET (PIP'5342 grant), Consiglio
Nazionale delle Ricerche (CNR Short-Term Mobility Grant 2006), as well as
hospitality and support from Dipartimento di Fisica (Univ. di Parma), 
Inst. f\"{u}r Theoretische Physik (Univ. zu K\"{o}ln) and ICTP.


\vfill\eject

\section{Appendix.}

\bigskip

\subsection{Evaluation of the AF order parameter in the hybrid $\left\{ A_{p%
\protect\sigma }^{\left( \dagger \right) },B_{p\protect\sigma }^{\left(
\dagger \right) }\right\} $ basis.}

Using the real-space representation we have to evaluate 
\begin{eqnarray}
s_{\mathcal{Q}}^{z}=\frac{1}{2N}\sum_{l\in A,\sigma }\sigma c_{l\sigma
}^{\dagger }c_{l\sigma }^{{}}-\frac{1}{2N}\sum_{j\in B\sigma }\sigma
c_{j\sigma }^{\dagger }c_{j\sigma }^{{}}
\label{staggered magn band in AB basis}
\end{eqnarray}

Notice that Eq.\ref{staggered magn band in AB basis} implicitly assumes that 
$\left\langle s_{\mathcal{Q}}^{z}\right\rangle $ is positive on the $%
\mathcal{A}$ sublattice, which is correct only for a FM\ Kondo coupling. To
keep track of the correct sign for arbitrary sign of $J_{K}$ we have to use
Eq.\ref{sign szband} yielding $\left\langle s_{\mathcal{Q}}^{z}\right\rangle
=-\mathrm{sgn}\left( J_{K}\right) \left\vert \left\langle s_{\mathcal{Q}%
}^{z}\right\rangle \right\vert $.

For a generic site $m\in \mathcal{A}\cup \mathcal{B}$ one has the standard
decomposition: 
\begin{eqnarray}
c_{m\sigma }^{\dagger }=\frac{1}{\sqrt{N}}\sum_{p\in RBZ}c_{p\sigma
}^{\dagger }e^{-ipR_{m}}+\frac{1}{\sqrt{N}}\sum_{p\in RBZ}c_{p+\mathbf{%
\mathcal{Q}}\sigma }^{\dagger }e^{-i(p+\mathbf{\mathcal{Q)}}R_{m}}
\end{eqnarray}

Substituting for \ $c_{p\sigma }^{\dagger },c_{p+\mathbf{\mathcal{Q}}\sigma
}^{\dagger }$\ \ \ the $\left\{ \alpha _{p\sigma }^{\left( \dagger \right)
},\beta _{p\sigma }^{\left( \dagger \right) }\right\} $operators and
recalling that on the $\mathcal{A}$ sites $R_{m}=2m$\textbf{$a$} while on $%
\mathcal{B}$ sites $R_{m}=\left( 2m+1\right) \mathbf{a}$ one has:

\begin{eqnarray}
R_{l} &\in &\mathcal{A}\qquad c_{l\sigma }^{\dagger }=\frac{1}{\sqrt{N}}%
\sum_{p\in RBZ}\left[ \left( \alpha _{p\sigma }^{\dagger }+\beta _{p\sigma
}^{\dagger }\right) \cos \zeta _{p}+\sigma \left( \alpha _{p\sigma
}^{\dagger }-\beta _{p\sigma }^{\dagger }\right) \sin \zeta _{p}\right]
e^{-ipR_{l}}  \notag \\
R_{j} &\in &\mathcal{B}\qquad c_{j\sigma }^{\dagger }=\frac{1}{\sqrt{N}}%
\sum_{p\in RBZ}\left[ \left( \alpha _{p\sigma }^{\dagger }-\beta _{p\sigma
}^{\dagger }\right) \cos \zeta _{p}-\sigma \left( \alpha _{p\sigma
}^{\dagger }+\beta _{p\sigma }^{\dagger }\right) \sin \zeta _{p}\right]
e^{-ipR_{j}}  \notag \\
&&
\end{eqnarray}

Expressing the $\left\{ \alpha _{p\sigma }^{\left( \dagger \right) },\beta
_{p\sigma }^{\left( \dagger \right) }\right\} $operators\ \ in terms of the$%
\left\{ A_{p\sigma }^{\left( \dagger \right) },B_{p\sigma }^{\left( \dagger
\right) }\right\} $basis the real space Fermi operators in the hybrid basis
for $R_{l}\in \mathcal{A}$ are given by: 
\begin{eqnarray}
c_{l\sigma }^{\dagger }=\frac{1}{\sqrt{N}}\sum_{p}\left[ \left( A_{p\sigma
}^{\dagger }+B_{p\sigma }^{\dagger }\right) \cos \left( \zeta _{p}-\xi
_{p}\right) +\sigma \left( A_{p\sigma }^{\dagger }-B_{p\sigma }^{\dagger
}\right) \sin \left( \zeta _{p}-\xi _{p}\right) \right] e^{-ipR_{l}}  \notag
\end{eqnarray}

and, for $R_{j}\in \mathcal{B}$, by: 
\begin{eqnarray}
c_{j\sigma }^{\dagger }=\frac{1}{\sqrt{N}}\sum_{p}\left[ \left( A_{p\sigma
}^{\dagger }-B_{p\sigma }^{\dagger }\right) \cos \left( \zeta _{p}+\xi
_{p}\right) -\sigma \left( A_{p\sigma }^{\dagger }+B_{p\sigma }^{\dagger
}\right) \sin \left( \zeta _{p}+\xi _{p}\right) \right] e^{-ipR_{j}}
\label{c_j,sigma^dagger for B subl in AB basis}
\end{eqnarray}

It follows:%
\begin{align}
\frac{1}{2N}\sum_{l\in \mathcal{A},\sigma }\sigma n_{l\sigma }^{{}}& =\frac{1%
}{4N}\sum_{p,\sigma }A_{p\sigma }^{\dagger }A_{p\sigma }^{{}}\left\{ \sigma
+\sin \left[ 2\left( \zeta _{p}-\xi _{p}\right) \right] \right\} +\frac{1}{4N%
}\sum_{p,\sigma }B_{p\sigma }^{\dagger }B_{\sigma }^{{}}\left\{ \sigma -\sin %
\left[ 2\left( \zeta _{p}-\xi _{p}\right) \right] \right\}  \notag \\
& +\frac{1}{4N}\sum_{p,\sigma }\left( A_{p\sigma }^{\dagger }B_{p\sigma
}^{{}}+B_{p\sigma }^{\dagger }A_{p\sigma }^{{}}\right) \sigma \cos \left[
2\left( \zeta _{p}-\xi _{p}\right) \right]  \label{band moment on A sites}
\end{align}%
and 
\begin{align}
-\frac{1}{2N}\sum_{j\in B\sigma }\sigma n_{j\sigma }^{{}}& =-\frac{1}{4N}%
\sum_{p}A_{p\sigma }^{\dagger }A_{p\sigma }^{{}}\left\{ \sigma -\sin \left[
2\left( \zeta _{p}+\xi _{p}\right) \right] \right\} -\frac{1}{4N}%
\sum_{p}B_{p\sigma }^{\dagger }B_{p\sigma }^{{}}\left\{ \sigma +\sin \left[
2\left( \zeta _{p}+\xi _{p}\right) \right] \right\}  \notag \\
& +\frac{1}{4N}\sum_{p}\sigma \cos \left[ 2\left( \zeta _{p}+\xi _{p}\right) %
\right] \left( A_{p\sigma }^{\dagger }B_{p\sigma }^{{}}+B_{p\sigma
}^{\dagger }A_{p\sigma }^{{}}\right)  \label{band moment on B sites}
\end{align}%
The Eqs.\ref{band moment on A sites} and \ref{band moment on B sites}, when
substituted into Eq.\ref{staggered magn band in AB basis}, yield:%
\begin{align}
s_{AB}^{z}& =\frac{1}{2N}\sum_{p,\sigma }\left\{ \sin \left[ 2\left( \zeta
_{p}+\xi _{p}\right) \right] A_{p\sigma }^{\dagger }A_{p\sigma }^{{}}-\sin %
\left[ 2\left( \zeta _{p}-\xi _{p}\right) \right] B_{p\sigma }^{\dagger
}B_{\sigma }^{{}}\right\}  \notag \\
& +\frac{1}{4N}\sum_{p,\sigma }\sigma \left( A_{p\sigma }^{\dagger
}B_{p\sigma }^{{}}+B_{p\sigma }^{\dagger }A_{p\sigma }^{{}}\right) \left\{
\cos \left[ 2\left( \zeta _{p}-\xi _{p}\right) \right] +\cos \left[ 2\left(
\zeta _{p}+\xi _{p}\right) \right] \right\}
\label{band AF moment in AB basis operator}
\end{align}%
When taking the Fermi average of Eq.\ref{band AF moment in AB basis operator}%
, the terms $A_{p\sigma }^{\dagger }B_{p\sigma }^{{}}+B_{p\sigma }^{\dagger
}A_{p\sigma }^{{}}$ do not contribute and we obtain Eq.\ref{band AF moment
in AB basis averaged}

\subsection{The coefficients $\mathcal{C}_{XY}^{\protect\lambda\protect\tau%
}\left( k,q\right) $ in Eq.\protect\ref{Kondo transverse a}.}

It is convenient to define for short 
\begin{align}
L_{k,p}^{\lambda} & =\cos\left( Z_{k}^{{}}+\lambda Z_{p}^{{}}\right) \qquad
M_{k,p}^{\lambda}=\sin\left( Z_{k}^{{}}+\lambda Z_{p}^{{}}\right)
\label{def.L_lambda M_lambda} \\
\qquad\lambda & =\pm\quad\quad Z_{p}=\zeta_{p}-\xi_{p}  \notag
\end{align}
notice that $L_{k,k}^{+}=\cos\left( 2Z_{k}^{{}}\right) $ and $%
M_{k,k}^{+}=\sin\left( 2Z_{k}^{{}}\right) $. In Eq.\ref{Kondo transverse a}
the coefficients $\mathcal{C}_{XY}^{+-}\left( k,q\right) $ and $\mathcal{C}%
_{XY}^{-+}\left( k,q\right) $ with $X,Y=A,B$ then read:%
\begin{align}
\mathcal{C}_{XY}^{+-}\left( k,q\right) & =\delta_{XY}\left[
L_{k,k+q}^{+}+\left( 2\delta_{XA}-1\right) M_{k,k+q}^{-}\right]  \notag \\
& +\left( 1-\delta_{XY}\right) \left[ L_{k,k+q}^{-}+\left( 1-2\delta
_{XB}\right) M_{k,k+q}^{+}\right]  \notag \\
\mathcal{C}_{XY}^{-+}\left( k,q\right) & =\delta_{XY}\mathcal{C}%
_{XX}^{+-}\left( k,q\right) -\left( 1-\delta_{XY}\right) \mathcal{C}%
_{YX}^{+-}\left( k,q\right)  \label{def.C_XY_pm}
\end{align}

\subsection{The generators $\mathcal{R}_{d}^{z}$ and \ $\mathcal{R}_{nd}^{z}$%
\ .}

In the perturbation we have several types of contributions. In $I_{d}^{z}$
and $I_{nd}^{z}$ we find terms with two types of products of Bose operators.
The generator corresponding to the perturbation term of first type
(number-conserving) like $\sum_{pqr,\sigma}\mathcal{C}_{pq,\sigma}X_{r,%
\sigma}^{\dagger}Y_{p-q+r,\sigma }^{{}}a_{p}^{\dagger}a_{q}^{{}}$ , where $%
X=A,B$ while $\mathcal{C}_{pq\sigma }$ is a numerical coefficient, is given
by: 
\begin{eqnarray}
\sum_{pqr,\sigma}\frac{\mathcal{C}_{pq\sigma}}{\mathcal{E}_{r,\sigma}^{X}-%
\mathcal{E}_{p-q+r,\sigma}^{Y}+\hbar\left( \Omega_{p}-\Omega_{q}\right) }%
X_{r\sigma}^{\dagger}Y_{p-q+r,\sigma}^{{}}a_{p}^{\dagger}a_{q}^{{}}
\end{eqnarray}

The generator corresponding to the perturbation term of the second type like 
$\sum_{pq,\sigma}\mathcal{C}_{pq,\sigma}X_{p,\sigma}^{\dagger}Y_{p,%
\sigma}^{{}}a_{q}^{\left( \dagger\right) }a_{-q}^{\left( \dagger\right) }$
reads%
\begin{eqnarray}
\sum_{pq,\sigma}\frac{\mathcal{C}_{pq\sigma}}{\mathcal{E}_{r,\sigma}^{X}-%
\mathcal{E}_{p,\sigma}^{Y}\pm2\hbar\Omega_{q}}X_{p,\sigma}^{\dagger
}Y_{p,\sigma}^{{}}a_{q}^{\left( \dagger\right) }a_{-q}^{\left(
\dagger\right) }
\end{eqnarray}
where the $\left( +\right) $ sign applies for bosonic creation operators,
and $\left( -\right) $ for destruction operators.

\subsection{The generator $\mathcal{R}_{{}}^{\perp}$ $.$}

The generator $\mathcal{R}_{{}}^{\perp}$ resulting from the transverse Kondo
term $I_{{}}^{\perp},$ Eq.\ref{Kondo transverse a}, can be written as the
sum of four contributions: $\mathcal{R}_{{}}^{\perp}=\sum_{X,Y=A,B}\mathcal{R%
}_{XY}^{\perp}$ , where:%
\begin{align}
\mathcal{R}_{XY}^{\perp} & = \frac{J_{K}}{2}\sqrt{\frac{S}{N}}\sum_{kq\sigma
}\left[ \delta_{XY}+\left( 1-\delta_{XY}\right) \sigma\right] \times  \notag
\\
& \times X_{k\sigma}^{\dagger}Y_{k+q,-\sigma}^{{}}\left( \mathcal{W}%
_{kq}^{XY}a_{q}^{\dagger}+\mathcal{Z}_{kq}^{XY}a_{-q}^{{}}\right)
\end{align}
and the coefficients $\mathcal{W}_{kq}^{XY}$ and $\mathcal{Z}_{kq}^{XY}$ are
given by: 
\begin{eqnarray}
\mathcal{W}_{kq}^{XY}=\frac{\left[ \mathrm{Ch}\left( \vartheta_{q}\right) 
\mathcal{C}_{XY}^{+-}\left( k,q\right) +\mathrm{Sh}\left( \vartheta
_{q}\right) \mathcal{C}_{XY}^{-+}\left( k,q\right) \right] }{\left( \mathcal{%
E}_{k}^{X}-\mathcal{E}_{k+q}^{Y}+\hbar\Omega_{q}^{{}}\right) }
\end{eqnarray}%
\begin{eqnarray}
\mathcal{Z}_{kq}^{XY}=\frac{\left[ \mathrm{Sh}\left( \vartheta_{q}\right) 
\mathcal{C}_{XY}^{+-}\left( k,q\right) +\mathrm{Ch}\left( \vartheta
_{q}\right) \mathcal{C}_{XY}^{-+}\left( k,q\right) \right] }{\left( \mathcal{%
E}_{k}^{X}-\mathcal{E}_{k+q}^{Y}-\hbar\Omega_{q}^{{}}\right) }
\end{eqnarray}

\subsection{The coefficients of $\left( 1/2\right) \left\langle \left[ 
\mathcal{R},I\right] \right\rangle _{Fermi}$ .}

We have obtained the effective Hamiltonian in Eq.\ref{H_eff_final}. Taking
advantage of the electron-hole symmetry, the coefficents will be now
explicitated assuming the paramagnetic band filling per site $n\leq1$ so
that $\left\langle n_{k,\sigma}^{B}\right\rangle =0$ in the ground state.

\subsubsection{The coefficients of $\left( 1/2\right) \left\langle \left[ 
\mathcal{R}_{d}^{z},I_{d}^{z}\right] \right\rangle _{Fermi}.$}

In Eq.\ref{Eff_Ham_long_diag} the coefficients read:

\begin{align}
\mathcal{G}_{q}^{har} & =-\frac{J_{K}^{2}}{16}\left( \frac{2}{N}\right)
^{2}\sum_{pr\sigma}M_{p,p}^{+}M_{r,r}^{+}\left[ \delta_{rp}+\left(
1-\delta_{rp}\right) \left\langle n_{r\sigma}^{A}\right\rangle \right]
\left\langle n_{p\sigma}^{A}\right\rangle \frac{\mathrm{Sh}^{2}\left(
2\vartheta_{q}\right) }{\hbar\Omega_{q}}  \label{def.L_har_q_nB0} \\
& +\frac{J_{K}^{2}}{8}\left( \frac{2}{N}\right) \sum_{p\sigma}\left(
L_{p,p}^{+}\right) ^{2}\left[ \frac{\mathrm{1}}{\mathcal{E}_{p\sigma}^{A}-%
\mathcal{E}_{p\sigma}^{B}}+\frac{\mathrm{1}}{\mathcal{E}_{p\sigma}^{A}-%
\mathcal{E}_{p\sigma}^{B}-2\hbar\Omega_{q}}\right] \left\langle
n_{p\sigma}^{A}\right\rangle \mathrm{Sh}^{2}\left( 2\vartheta_{q}\right) 
\notag \\
& +\frac{J_{K}^{2}}{8}\left( \frac{2}{N}\right) \sum_{p\sigma}\left(
L_{p,p}^{+}\right) ^{2}\frac{\left\langle n_{p\sigma}^{A}\right\rangle }{%
\mathcal{E}_{p\sigma}^{A}-\mathcal{E}_{p\sigma}^{B}}  \notag
\end{align}

and 
\begin{align}
\ \mathcal{G}_{q}^{anhar} & =-\frac{J_{K}^{2}}{16}\left( \frac{2}{N}\right)
^{2}\sum_{pr\sigma}M_{p,p}^{+}M_{r,r}^{+}\left[ \delta _{rp}+\left(
1-\delta_{rp}\right) \left\langle n_{r\sigma}^{A}\right\rangle \right]
\left\langle n_{p\sigma}^{A}\right\rangle \frac{\mathrm{Sh}\left(
4\vartheta_{q}\right) }{2\hbar\Omega_{q}}  \label{def.L_anhar_q_nB0} \\
& +\frac{J_{K}^{2}}{8}\left( \frac{2}{N}\right) \sum_{pq\sigma}\left(
L_{p,p}^{+}\right) ^{2}\left[ \frac{1}{\mathcal{E}_{p\sigma}^{A}-\mathcal{E}%
_{p\sigma}^{B}}+\frac{1}{\mathcal{E}_{p\sigma}^{A}-\mathcal{E}%
_{p\sigma}^{B}-2\hbar\Omega_{q}}\right] \left\langle
n_{p\sigma}^{A}\right\rangle \frac{\mathrm{Sh}\left( 4\vartheta_{q}\right) }{%
2}  \notag
\end{align}

\subsubsection{The coefficients of $\left( 1/2\right) \left\langle \left[ 
\mathcal{R}_{nd}^{z},I_{nd}^{z}\right] \right\rangle _{Fermi}.$}

To write down the coefficients $d_{q}^{z\pm}$ and $\varpi_{q}^{z\pm}$ of Eq.%
\ref{Eff_Ham_long_non_diag} it is convenient to introduce:%
\begin{eqnarray}
\mathrm{Ch}\left( \vartheta_{q}+\vartheta_{p}\right) =\mathfrak{C}_{qp}\quad%
\mathrm{Sh}\left( \vartheta_{q}+\vartheta_{p}\right) =\mathfrak{S}_{qp}
\end{eqnarray}
By defining 
\begin{align}
\mathcal{L}_{pqr}^{+} & =-M_{r,p-q+r}^{+}\left( \mathcal{X}%
_{q,p-q+r,p}^{AA1}+\mathcal{X}_{-p,p-q+r,-q}^{AA2}\right) \mathfrak{C}%
_{qp}\left\langle n_{p-q+r,\sigma}^{A}\right\rangle  \notag \\
& -L_{r,p-q+r}^{+}\left( \mathcal{X}_{q,p-q+r,p}^{AB1}+\mathcal{X}%
_{-p,p-q+r,-q}^{AB2}\right) \mathfrak{C}_{qp}\left\langle n_{p-q+r,\sigma
}^{A}\right\rangle  \notag \\
& +M_{r,p-q+r}^{+}\left( \mathcal{X}_{q,p-q+r,p}^{AA3}+\mathcal{X}%
_{-p,p-q+r,-q}^{AA3}\right) \mathfrak{S}_{qp}\left\langle
n_{r\sigma}^{A}\right\rangle  \notag \\
& +L_{r,p-q+r}^{+}\left( \mathcal{X}_{q,p-q+r,p}^{BA3}+\mathcal{X}%
_{-p,p-q+r,-q}^{BA3}\right) \mathfrak{S}_{qp}\left\langle
n_{r\sigma}^{A}\right\rangle  \notag \\
&
\end{align}
and 
\begin{align}
\mathcal{L}_{pqr}^{-} & =+M_{r,p-q+r}^{+}\left( \mathcal{X}%
_{-p,p-q+r,-q}^{AA1}+\widetilde{\mathcal{X}}_{q,p-q+r,p}^{AA2}\right) 
\mathfrak{C}_{qp}\left\langle n_{r\sigma}^{A}\right\rangle  \notag \\
& +L_{r,p-q+r}^{+}\left( \mathcal{X}_{-p,p-q+r,-q}^{BA1}+\mathcal{X}%
_{q,p-q+r,p}^{BA2}\right) \mathfrak{C}_{qp}\left\langle
n_{r\sigma}^{A}\right\rangle  \notag \\
& -M_{r,p-q+r}^{+}\left( \mathcal{X}_{q,p-q+r,p}^{AA4}+\mathcal{X}%
_{-p,p-q+r,-q}^{AA4}\right) \mathfrak{S}_{qp}\left\langle n_{p-q+r,\sigma
}^{A}\right\rangle  \notag \\
& -L_{r,p-q+r}^{+}\left( \mathcal{X}_{q,p-q+r,p}^{AB4}+\mathcal{X}%
_{-p,p-q+r,-q}^{AB4}\right) \mathfrak{S}_{qp}\left\langle n_{p-q+r,\sigma
}^{A}\right\rangle  \notag \\
&
\end{align}
where ($X=A,B$\ ) 
\begin{eqnarray}
\mathcal{X}_{xyw}^{XX1}=-\frac{\left( 2\delta_{XA}-1\right) \sin\left(
Z_{y}+Z_{x-w+y}\right) }{\left[ \mathcal{E}_{y}^{X}-\mathcal{E}%
_{x+y-w}^{X}+\hbar\left( \Omega_{w}-\Omega_{x}\right) \right] }\mathrm{Ch}%
\left( \vartheta_{w}\right) \mathrm{Ch}\left( \vartheta_{x}\right)
\end{eqnarray}%
\begin{eqnarray}
\mathcal{X}_{xyw}^{XX2}=-\frac{\left( 2\delta_{XA}-1\right) \sin\left(
Z_{y}+Z_{x-w+y}\right) }{\left[ \mathcal{E}_{y}^{X}-\mathcal{E}%
_{x+y-w}^{X}+\hbar\left( \Omega_{w}-\Omega_{x}\right) \right] }\mathrm{Sh}%
\left( \vartheta_{w}\right) \mathrm{Sh}\left( \vartheta_{x}\right)
\end{eqnarray}%
\begin{eqnarray}
\mathcal{X}_{xyw}^{XX3}=-\frac{\left( 2\delta_{XA}-1\right) \sin\left(
Z_{y}+Z_{x-w+y}\right) }{\left[ \mathcal{E}_{y}^{X}-\mathcal{E}%
_{x+y-w}^{X}+\hbar\left( \Omega_{w}+\Omega_{x}\right) \right] }\mathrm{Sh}%
\left( \vartheta_{w}\right) \mathrm{Ch}\left( \vartheta_{x}\right)
\end{eqnarray}%
\begin{eqnarray}
\mathcal{X}_{xyw}^{XX4}=-\frac{\left( 2\delta_{XA}-1\right) \sin\left(
Z_{y}+Z_{x-w+y}\right) }{\left[ \mathcal{E}_{y}^{X}-\mathcal{E}%
_{x+y-w}^{X}-\hbar\left( \Omega_{w}+\Omega_{x}\right) \right] }\mathrm{Sh}%
\left( \vartheta_{w}\right) \mathrm{Ch}\left( \vartheta_{x}\right)
\end{eqnarray}%
\begin{eqnarray}
\mathcal{X}_{xyw}^{XY1}=-\frac{\cos\left( Z_{y}+Z_{x-w+y}\right) }{\left[ 
\mathcal{E}_{y}^{X}-\mathcal{E}_{x+y-w}^{Y}+\hbar\left( \Omega_{w}-\Omega
_{x}\right) \right] }\mathrm{Ch}\left( \vartheta_{w}\right) \mathrm{Ch}%
\left( \vartheta_{x}\right)
\end{eqnarray}%
\begin{eqnarray}
\mathcal{X}_{xyw}^{XY2}=-\frac{\cos\left( Z_{y}+Z_{x-w+y}\right) }{\left[ 
\mathcal{E}_{y}^{X}-\mathcal{E}_{x+y-w}^{Y}+\hbar\left( \Omega_{w}-\Omega
_{x}\right) \right] }\mathrm{Sh}\left( \vartheta_{w}\right) \mathrm{Sh}%
\left( \vartheta_{x}\right)
\end{eqnarray}%
\begin{eqnarray}
\mathcal{X}_{xyw}^{XY3}=-\frac{\cos\left( Z_{y}+Z_{x-w+y}\right) }{\left[ 
\mathcal{E}_{y}^{X}-\mathcal{E}_{x+y-w}^{Y}+\hbar\left( \Omega_{w}+\Omega
_{x}\right) \right] }\mathrm{Sh}\left( \vartheta_{w}\right) \mathrm{Ch}%
\left( \vartheta_{x}\right)
\end{eqnarray}%
\begin{eqnarray}
\mathcal{X}_{xyw}^{XY4}=-\frac{\cos\left( Z_{y}+Z_{x-w+y}\right) }{\left[ 
\mathcal{E}_{y}^{X}-\mathcal{E}_{x+y-w}^{Y}-\hbar\left( \Omega_{w}+\Omega
_{x}\right) \right] }\mathrm{Sh}\left( \vartheta_{w}\right) \mathrm{Ch}%
\left( \vartheta_{x}\right)
\end{eqnarray}
we can write 
\begin{eqnarray}
\hbar d_{q}^{z\pm}=\frac{J_{K}^{2}}{2}\left( \frac{2}{N}\right)
^{2}\sum_{pr}\left( 1-\delta_{pq}\right) \mathcal{L}_{pqr}^{\pm}
\end{eqnarray}

Next, by defining%
\begin{align}
\mathcal{M}_{pqr}^{+} & =-M_{r,p-q+r}^{+}\left( \mathcal{X}%
_{q,p-q+r,p}^{AA1}+\mathcal{X}_{-p,p-q+r,-q}^{AA2}\right) \mathfrak{S}%
_{qp}\left\langle n_{p-q+r,\sigma}^{A}\right\rangle  \notag \\
& -L_{r,p-q+r}^{+}\left( \mathcal{X}_{q,p-q+r,p}^{AB1}+\mathcal{X}%
_{-p,p-q+r,-q}^{AB2}\right) \mathfrak{S}_{qp}\left\langle n_{p-q+r,\sigma
}^{A}\right\rangle  \notag \\
& +M_{r,p-q+r}^{+}\left( \mathcal{X}_{q,p-q+r,p}^{AA3}+\mathcal{X}%
_{-p,p-q+r,-q}^{AA3}\right) \mathfrak{C}_{qp}\left\langle
n_{r\sigma}^{A}\right\rangle  \notag \\
& +L_{r,p-q+r}^{+}\left( \mathcal{X}_{q,p-q+r,p}^{BA3}+\mathcal{X}%
_{-p,p-q+r,-q}^{BA3}\right) \mathfrak{C}_{qp}\left\langle
n_{r\sigma}^{A}\right\rangle  \notag \\
&
\end{align}

and 
\begin{align}
\mathcal{M}_{pqr}^{-} & =+M_{r,p-q+r}^{+}\left( \mathcal{X}%
_{-p,p-q+r.-q}^{AA1}+\mathcal{X}_{q,p-q+r,p}^{AA2}\right) \mathfrak{S}%
_{qp}\left\langle n_{r\sigma}^{A}\right\rangle  \notag \\
& +L_{r,p-q+r}^{+}\left( \mathcal{X}_{-p,p-q+r,-q}^{BA1}+\mathcal{X}%
_{q,p-q+r,p}^{BA2}\right) \mathfrak{S}_{qp}\left\langle
n_{r\sigma}^{A}\right\rangle  \notag \\
& -M_{r,p-q+r}^{+}\left( \mathcal{X}_{q,p-q+r,p}^{AA4}+\mathcal{X}%
_{-p,p-q+r,-q}^{AA4}\right) \mathfrak{C}_{qp}\left\langle n_{p-q+r,\sigma
}^{A}\right\rangle  \notag \\
& -L_{r,p-q+r}^{+}\left( \mathcal{X}_{q,p-q+r,p}^{AB4}+\mathcal{X}%
_{-p,p-q+r,-q}^{AB4}\right) \mathfrak{C}_{qp}\left\langle n_{p-q+r,\sigma
}^{A}\right\rangle  \notag \\
&
\end{align}

we can write 
\begin{eqnarray}
\hbar\mathcal{\varpi}_{q}^{z\pm}=\frac{J_{K}^{2}}{2}\left( \frac{2}{N}%
\right) ^{2}\sum_{pr}\left( 1-\delta_{pq}\right) \mathcal{M}_{pqr}^{\pm}
\label{def.omega-small_z_pm}
\end{eqnarray}

\subsubsection{The coefficients of $\left( 1/2\right) \left\langle \left[ 
\mathcal{R}^{\perp},I_{{}}^{\perp}\right] \right\rangle _{Fermi}.$}

In Eq.\ref{Eff_Ham_trans_non_diag} , if $n\leq1$ , so that $\left\langle
n_{k,\sigma}^{B}\right\rangle =0$ in the ground state, one finds $\ \ 
\mathcal{T}_{q}^{BB}=\mathcal{S}_{q}^{BB}=0.$ The other $\mathcal{T}%
_{q}^{XY} $ coefficients, by defining $\left\langle f_{k,k\pm
q,\sigma}^{A}\right\rangle =\left\langle n_{k,-\sigma}^{A}\right\rangle
-\left\langle n_{k\pm q,\sigma}^{A}\right\rangle $ , read:

\begin{align}
\mathcal{T}_{q}^{AA} & =\frac{J_{K}^{2}S}{4N}\sum_{k\sigma}\mathcal{W}%
_{k,q}^{AA}\mathrm{Sh}\left( \vartheta_{q}\right) \mathcal{C}%
_{AA}^{+-}\left( k+q,-q\right) \left\langle f_{k,k+q,\sigma}^{A}\right\rangle
\notag \\
& +\frac{J_{K}^{2}S}{4N}\sum_{k\sigma}\mathcal{W}_{k,q}^{AA}\mathrm{Ch}%
\left( \vartheta_{q}\right) \mathcal{C}_{BB}^{+-}\left( k+q,-q\right)
\left\langle f_{k,k+q,\sigma}^{A}\right\rangle  \notag \\
& +\frac{J_{K}^{2}S}{4N}\sum_{k\sigma}\mathcal{Z}_{k,-q}^{AA}\mathrm{Ch}%
\left( \vartheta_{q}\right) \mathcal{C}_{AA}^{+-}\left( k-q,q\right)
\left\langle f_{k,k-q,\sigma}^{A}\right\rangle  \notag \\
& +\frac{J_{K}^{2}S}{4N}\sum_{k\sigma}\mathcal{Z}_{k,-q}^{AA}\mathrm{Sh}%
\left( \vartheta_{q}\right) \mathcal{C}_{BB}^{+-}\left( k-q,q\right)
\left\langle f_{k,k-q,\sigma}^{A}\right\rangle
\end{align}

\begin{align}
\mathcal{T}_{q}^{AB} & =-\frac{J_{K}^{2}S}{2N}\sum_{k\sigma}\mathcal{W}%
_{k,q}^{AB}\mathrm{Sh}\left( \vartheta_{q}\right) \mathcal{C}%
_{BA}^{+-}\left( k+q,-q\right) \left\langle n_{k,-\sigma}^{A}\right\rangle 
\notag \\
& +\frac{J_{K}^{2}S}{2N}\sum_{k\sigma}\mathcal{W}_{k,q}^{AB}\mathrm{Ch}%
\left( \vartheta_{q}\right) \mathcal{C}_{AB}^{+-}\left( k+q,-q\right)
\left\langle n_{k,-\sigma}^{A}\right\rangle  \notag \\
& -\frac{J_{K}^{2}S}{2N}\sum_{k\sigma}\mathcal{Z}_{k,-q}^{AB}\mathrm{Ch}%
\left( \vartheta_{q}\right) \mathcal{C}_{BA}^{+-}\left( k-q,q\right)
\left\langle n_{k,-\sigma}^{A}\right\rangle  \notag \\
& +\frac{J_{K}^{2}S}{2N}\sum_{k\sigma}\mathcal{Z}_{k,-q}^{AB}\mathrm{Sh}%
\left( \vartheta_{q}\right) \mathcal{C}_{AB}^{+-}\left( k-q,q\right)
\left\langle n_{k,-\sigma}^{A}\right\rangle
\end{align}
and 
\begin{align}
\mathcal{T}_{q}^{BA} & =+\frac{J_{K}^{2}S}{2N}\sum_{k}\mathcal{W}_{k,q}^{BA}%
\mathrm{Sh}\left( \vartheta_{q}\right) \mathcal{C}_{AB}^{+-}\left(
k+q,-q\right) \left\langle n_{k+q,\sigma}^{A}\right\rangle  \notag \\
& -\frac{J_{K}^{2}S}{2N}\sum_{k}\mathcal{W}_{k,q}^{BA}\mathrm{Ch}\left(
\vartheta_{q}\right) \mathcal{C}_{BA}^{+-}\left( k+q,-q\right) \left\langle
n_{k+q,\sigma}^{A}\right\rangle  \notag \\
& +\frac{J_{K}^{2}S}{2N}\sum_{k}\mathcal{Z}_{k,-q}^{BA}\mathrm{Ch}\left(
\vartheta_{q}\right) \mathcal{C}_{AB}^{+-}\left( k-q,q\right) \left\langle
n_{k-q,\sigma}^{A}\right\rangle  \notag \\
& -\frac{J_{K}^{2}S}{2N}\sum_{k}\mathcal{Z}_{k,-q}^{BA}\mathrm{Sh}\left(
\vartheta_{q}\right) \mathcal{C}_{BA}^{+-}\left( k-q,q\right) \left\langle
n_{k-q,\sigma}^{A}\right\rangle  \notag \\
&
\end{align}

The $\mathcal{S}_{q}^{XY1}$ coefficients read:%
\begin{align}
\mathcal{S}_{q}^{AA1} & =+\frac{J_{K}^{2}S}{2N}\sum_{k}\mathcal{W}%
_{k,-q}^{AA}\mathrm{Ch}\left( \vartheta_{q}\right) \mathcal{C}%
_{AA}^{+-}\left( k-q,q\right) \left\langle N_{k,-q,\sigma}\right\rangle 
\notag \\
& +\frac{J_{K}^{2}S}{2N}\sum_{k}\mathcal{W}_{k,-q}^{AA}\mathrm{Sh}\left(
\vartheta_{q}\right) \mathcal{C}_{BB}^{+-}\left( k-q,q\right) \left\langle
N_{k,-q,\sigma}\right\rangle  \notag \\
&
\end{align}%
\begin{align}
\mathcal{S}_{q}^{AB1} & =-\frac{J_{K}^{2}S}{2N}\sum_{k}\mathcal{W}%
_{k,-q}^{AB}\mathrm{Ch}\left( \vartheta_{q}\right) \mathcal{C}%
_{BA}^{+-}\left( k-q,q\right) \left\langle n_{k,-\sigma}^{A}\right\rangle 
\notag \\
& +\frac{J_{K}^{2}S}{2N}\sum_{k}\mathcal{W}_{k,-q}^{AB}\mathrm{Sh}\left(
\vartheta_{q}\right) \mathcal{C}_{AB}^{+-}\left( k-q,q\right) \left\langle
n_{k,-\sigma}^{A}\right\rangle  \notag \\
&
\end{align}%
\begin{align}
\mathcal{S}_{q}^{BA1} & =+\frac{J_{K}^{2}S}{2N}\sum_{k}\mathcal{W}%
_{k,-q}^{BA}\mathrm{Ch}\left( \vartheta_{q}\right) \mathcal{C}%
_{AB}^{+-}\left( k-q,q\right) \left\langle n_{k-q,\sigma}^{A}\right\rangle 
\notag \\
& -\frac{J_{K}^{2}S}{2N}\sum_{k}\mathcal{W}_{k,-q}^{BA}\mathrm{Sh}\left(
\vartheta_{q}\right) \mathcal{C}_{BA}^{+-}\left( k-q,q\right) \left\langle
n_{k-q,\sigma}^{A}\right\rangle  \notag \\
&
\end{align}

The coefficients $\mathcal{S}_{q}^{XY2}$ can be obtained from
$\mathcal{S}_{q}^{XY1}$ by interchanging $\mathcal{W}_{k,-q}^{XY}$ 
and $\mathrm{Ch} \left( \vartheta _{q}\right) $ 
respectively with $\mathcal{Z}_{k,-q}^{XY}$
and $\mathrm{Sh}\left( \vartheta _{q}\right)$ .




%
%

%
%
%
%

\newpage


\begin{table}[tbp]
\caption{Conduction band AF spin polarization, $<s^{z}>_{AB}$, dependence on other model
parameters, for the cases depicted in respective Figs. 3 to 6. In first
column we enter the relative Kondo coupling magnitude (cases of Fig. 3:
notice that $J_K/W \sim J_K/J_H \times 10^{-4}$, here), while in second
column we state the corresponding band polarization $<s^{z}>_{AB}$ obtained; similarly,
for Fig.4 and the related 3rd and 4th columns of this table. 
In 5th column we enter the electron correlation $U/t$ cases from Fig. 5, and
next column states the corresponding $<s^{z}>_{AB}$. In 7th column we enter filling $n$ values of
Fig. 6, and in last column the corresponding $<s^{z}>_{AB}$ values. }

\bigskip

\label{table1}
\begin{tabular}{cccccccc}
\multicolumn{1}{c}{$J_K/J_H$ (PM)} & \multicolumn{1}{c}{$<s^{z}>_{AB}$} &
\multicolumn{1}{c}{$J_K/J_H$} (AF) & \multicolumn{1}{c}{$<s^{z}>_{AB}$} &
\multicolumn{1}{c}{$\qquad U/t$} & \multicolumn{1}{c}{$<s^{z}>_{AB}$} &
\multicolumn{1}{c}{$\qquad n$} & \multicolumn{1}{c}{$<s^{z}>_{AB}$} \\


\hline  0.1 & -6.10E-7 & 1.0 & -7.20E-2 & 0.100 & -1.14E-4 & 0.100 & -3.81E-6 \\
\ 1.0 & -6.10E-6 & 4.0 & -7.20E-2 & 1.000 & -1.15E-4 & 0.500 & -2.96E-5 \\
\ 5.0 & -3.0E-5 & 5.5 & -7.20E-2 & 3.000 & -1.22E-4 & 0.600 & -3.97E-5 \\
10.0 & -6.10E-5 & 7.0 & -7.20E-2 & 5.000 & -1.42E-4 & 0.700 & -5.24E-5 \\
15.0 & -9.15E-5 &  &  & 5.500 & -1.51E-4 & 0.800 & -6.92E-5 \\
20.0 & -1.22E-4 &  &  & 5.800 & -3.52E-3 & 0.850 & -8.00E-5 \\
25.0 & -1.50E-4 &  &  & 5.8650 & -8.69E-3 & 0.900 & -9.94E-5 \\
30.0 & -1.80E-4 &  &  & 5.8652 & -8.71E-3 & 0.999 & -8.69E-3 

\end{tabular}
\end{table}


\end{document}